\documentclass[12pt,letterpaper]{article}

\usepackage[utf8]{inputenc}

\usepackage{latexsym}
\usepackage{verbatim}
\usepackage[a4paper, total={6.5in, 9in}]{geometry}
\usepackage{silence}
\usepackage{amsthm,amsmath,amsfonts,amssymb,verbatim, color}
\usepackage{graphicx}
\usepackage[percent]{overpic}
\usepackage{bm}
\usepackage{braket}
\usepackage{float}
\usepackage{mathtools}
\usepackage{appendix}
\usepackage[colorlinks=true,citecolor=blue,linkcolor=black,urlcolor=blue]{hyperref}
\usepackage{graphics} 
\graphicspath{ {figures/} }
\usepackage[caption=false]{subfig}

\usepackage[mathscr]{euscript}
\usepackage[export]{adjustbox}
\usepackage[bottom]{footmisc}
\usepackage{textcomp}
\usepackage{braket}

\title{Quantum Zeno effect: a qutrit controlled by a qubit}
\author{$^{1,2}$Komal Kumari, $^{1,2}$Garima Rajpoot, and $^{3}$Sudhir Ranjan Jain \\ $^{1}${\small{\it Theoretical Nuclear Physics and Quantum Computing Section}}\\ {\small{\it Nuclear Physics Division, Bhabha Atomic Research Centre, Mumbai 400085, India}}
\\
$^{2}${\small{\it Homi Bhabha National Institute, New Training School Complex, Anushakti Nagar, Mumbai 400094}}
\\$^{3}${\small{\it UM-DAE Centre for Excellence in Basic Sciences, Vidyanagari Campus}} \\ \small{\it University of Mumbai, Mumbai 400098, India}}
\date{March 2023}

\begin{document}

\maketitle
\begin{abstract}
For a three-level system monitored by an ancilla, we show that quantum Zeno effect can be employed to control quantum jump for error correction. Further, we show that we can realize cNOT gate, and effect dense coding and teleportation. We believe that this work paves the way to generalize the control of a qudit. 
\end{abstract}
\section{Introduction}
Quantum errors can be corrected only by developing methods to control quantum jumps. Recently, the quantum Zeno effect \cite{ms} has been employed to delay spontaneous emission, giving us time to detect possible erroneous jumps. Moreover, to observe and hence control quantum jumps, QZE has been shown to realize Dehmelt-like shelving \cite{dehmelt,deh}. This work was inspired by a very interesting and important experiment on ``catching" and ``reversing" a quantum jump by Minev et al. \cite{minev}. To take these thoughts further for realistic applications, we need to show this method of control for multi-level systems. Here we take the next step and consider a three-level system which has the possibility of three distinct frequencies $\omega_{12}$, $\omega_{23}$ and $\omega_{13}$. One of these states is monitored by a detector: a two-level ancillary qubit \cite{parveen,krjj}. In contrast to the control of two-level system where there is just one frequency, here there are three frequencies. Thus there are multiple time-scales under consideration. The aim of this article is to study the possibility of controlling spontaneous errors and shelving in the sense of Dehmelt and improvised in \cite{minev,krjj}.

The plan of the paper is as follows. In Section 2.1, we state the problem and present the principle of least action approach relevant to our physical situation. This is based on the mathematical treatment of $n-$ level system, the details of which are reviewed in the Appendix. The solution of the evolution equation of the density matrix in terms of coordinates and conjugate momenta is shown. In Section 2.2, the construction of a cNOT gate using a three-level system is explained.  It is interesting to see that the three-level system considered here can be related to dense coding and teleportation, explained in Sections 2.3 and 2.4.  

\section{Qutrit dynamics}

We have a three-level system, i.e., a qutrit, with levels $\ket{1}$, $\ket{2}$ and $\ket{3}$ and transition frequencies $\omega_{12}$, $\omega_{23}$ and $\omega_{31}$.
\begin{figure*}[ht!]
    \centering
    \subfloat[]{\includegraphics[width=0.5\textwidth]{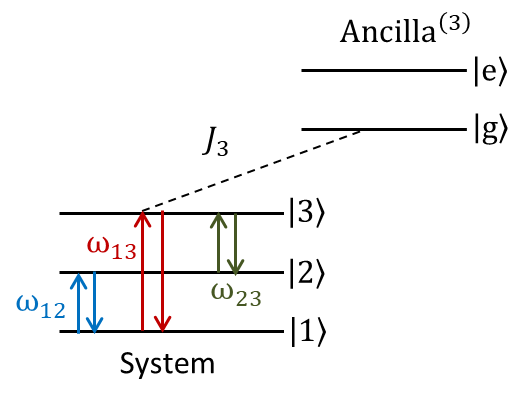}} 
    \caption{A qutrit is interacting with an ancilla or detector. Ancilla$^{(3)}$ is a two-level system which monitors the state $\ket{3}$ of the qutrit with a coupling strength $J_3$. The transition frequencies of the three-levels are $\omega_{12}$ between states $\ket{1}$ and $\ket{2}$, $\omega_{23}$ between states $\ket{2}$ and $\ket{3}$ and $\omega_{13}$ between states $\ket{1}$ and $\ket{3}$.}
\label{fig:system-single}
\end{figure*}
For a three-level system, $N=3$, the density matrix is 
\begin{equation}
    \rho=\frac{1}{3}\hat{\mathbb{I}}+\frac{1}{2}\sum_{i=1}^8x_i\hat{x}_i,
\end{equation}
where $1\leq j<k\leq N$, $1\leq l\leq N-1$ \cite{kimura}. For a detailed description, see Appendix. The operators are
\begin{alignat}{1}
    \hat{x}_1 &= \hat{u}_{12} = \ket{1}\bra{2}+\ket{2}\bra{1}\nonumber\\
    \hat{x}_2 &= \hat{v}_{12} = -\iota(\ket{1}\bra{2}-\ket{2}\bra{1})\nonumber\\
    \hat{x}_3 &= \hat{w}_{1} = \ket{1}\bra{1}-\ket{2}\bra{2}\nonumber\\
    \hat{x}_4 &= \hat{u}_{13} = \ket{1}\bra{3}+\ket{3}\bra{1}\nonumber\\
    \hat{x}_5 &= \hat{v}_{13} = -\iota(\ket{1}\bra{3}-\ket{3}\bra{1})\nonumber\\
    \hat{x}_6 &= \hat{u}_{23} = \ket{2}\bra{3}+\ket{3}\bra{2}\nonumber\\
    \hat{x}_7 &= \hat{v}_{23} = -\iota(\ket{2}\bra{3}-\ket{3}\bra{2})\nonumber\\
    \hat{x}_8 &= \hat{w}_{2} = \sqrt{\frac{1}{3}}(\ket{1}\bra{1}+\ket{2}\bra{2}-2\ket{3}\bra{3}).
\end{alignat}
The density operator in the matrix form is
\begin{alignat}{1}
    \hat{\rho}&=\begin{bmatrix}
        \frac{1}{3}+\frac{x_3}{2}+\frac{x_8}{\sqrt{3}} & \frac{1}{2}(x_1-\iota x_2) & \frac{1}{2}(x_4-\iota x_5)\\
        \frac{1}{2}(x_1+\iota x_2) & \frac{1}{3}-\frac{x_3}{2}+\frac{x_8}{\sqrt{3}} & \frac{1}{2}(x_6-\iota x_7)\\
        \frac{1}{2}(x_4+\iota x_5) & \frac{1}{2}(x_6+\iota x_7) & \frac{1}{3}-\frac{2x_8}{\sqrt{3}}
    \end{bmatrix}.
\end{alignat}

\subsection{Monitoring a single level}
Consider that the qutrit is interacting with an ancilla, a two-level system prepared initially in the state $\ket{0}$ of $\sigma_z$, Fig. \ref{fig:system-single}. The ancilla monitors the third level of the qutrit with a coupling strength $J_3=\sqrt{\frac{\alpha_3}{\delta t}}$, where $\alpha_3$ is a stochastic parameter related to the frequency of the detector. The qutrit+ancilla system evolves for a time $\delta t$ and then its $\sigma_y$ operator is measured. If the outcome of measurement is $0$, qutrit is in state $\ket{1}$ or $\ket{2}$.  This evolution and measurement is performed $n$ times for a total time of $T=n\delta t$. The ancilla is reset after every measurement. The Hamiltonian of the qutrit+ancilla system is 
\begin{alignat}{1}
    H&=H_s+H_{s-d}\nonumber\\
    &=\omega_{12}(\ket{1}\bra{2}+\ket{2}\bra{1}) + \omega_{23}(\ket{2}\bra{3}+\ket{3}\bra{2})+ \omega_{13}(\ket{1}\bra{3}+\ket{3}\bra{1})+J \ket{3}\bra{3}\otimes \sigma_y^{(3)},
\end{alignat}
where $H_{s-d}=J\ket{3}\bra{3}\otimes\sigma_y^{(3)}$, denoting that the state $\ket{3}$ is entangled with the ancilla and a measurement of the $y$ observable of the ancilla. The Kraus operators for measurement are given by
\begin{alignat}{1}
    \mathcal{M}_r&=\bra{r}\exp{[-\iota H_{s-d}\delta t]}\ket{0}\nonumber\\
    &= \bra{r}\mathbb{I}-\iota H_{s-d} \delta t -\frac{1}{2}H_{s-d}^2 (\delta t)^2\ket{0}\nonumber\\
    \mathcal{M}_0 &= \mathbb{I}-\frac{\alpha_3}{2}\ket{3}\bra{3}\delta t\\
    \mathcal{M}_1 &=\sqrt{\alpha_3\delta t}\ket{3}\bra{3}.
\end{alignat}

\begin{figure*}[ht!]
    \centering
    \subfloat[]{\includegraphics[width=0.35\textwidth]{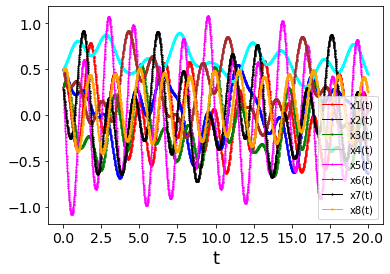}} 
    \subfloat[]{\includegraphics[width=0.35\textwidth]{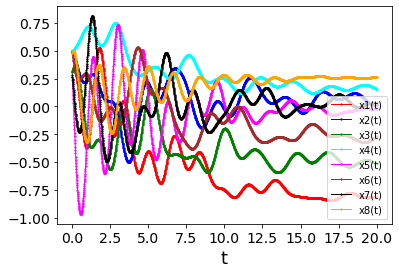}} 
    \subfloat[]{\includegraphics[width=0.35\textwidth]{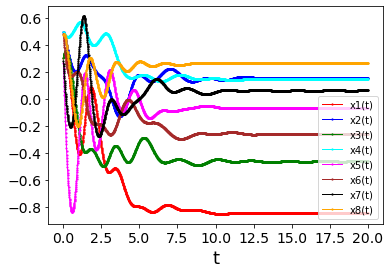}}
    \caption{Figure shows the dynamics of the three-level system, the variation of its 8 variables plotted with time when the third level is being monitored. The initial conditions have been chosen as $x_1=x_3=x_5=x_7=0.3$, $x_2=x_4=x_6=0.5$ and $x_8=\sqrt{(4/3)^2-(\sum_i x_i)^2}$, where $i=1,2,\dots,7$. The Rabi frequencies of the three levels is chosen to be $\omega_{12}=0.6$, $\omega_{23}=1$ and $\omega_{13}=1.6$. The system is being monitored in three frequency ranges, (a) $\alpha_3=0.2$, (b) $\alpha_3=0.7$ and (c) $\alpha_3=1.7$. In fig. (a), there are usual coherent oscillation. In (b), the system begins to freeze fairly early at a particular state. In fig. (c) the Zeno regime has set in.}
\label{fig:single_dynamics}
\end{figure*}

Upon unitary evolution of system via the operator $\mathcal{U}=\exp{-\iota H_s\delta t}$ and measurements post-selected on $t=0$, we obtain 
\begin{equation}
    \rho(t+\delta t)=\frac{\mathcal{M}^0 \mathcal{U}\rho\mathcal{U}^\dagger\mathcal{M}^{0\dagger}}{Tr[\mathcal{M}^0 \mathcal{U}\rho\mathcal{U}^\dagger\mathcal{M}^{0\dagger}]}.
\end{equation}
By extremising the action obtained for the Joint Probability Distribution Function (JPDF) for the system, we obtain eight coupled equations, their canonical conjugates, and a functional $\mathcal{F}$ incorporating the back-action of measurement performed by the detector \cite{krjj,jordan, jordan2}
\begin{alignat}{1}
    \dot{x}_1&=\omega_{23}x_5+\omega_{13}x_7+\frac{1}{3}\alpha_3x_1(1-2\sqrt{3}x_8)\nonumber\\
    \dot{x}_2&=-2\omega_{12}x_3-\omega_{23}x_4+\omega_{13}x_6+\frac{\alpha_3}{3}x_2(1-2\sqrt{3}x_8)\nonumber\\
    \dot{x}_3&=2\omega_{12}x_2+\omega_{13}x_5-\omega_{23}x_7+\frac{\alpha_3}{3}x_3(1-2\sqrt{3}x_8)\nonumber\\
    \dot{x}_4&= \omega_{23}x_2-\omega_{12}x_7-\frac{\alpha_3}{6}x_4(1+4\sqrt{3}x_8)\nonumber\\
    \dot{x}_5 &= -\omega_{23}x_1 +\omega_{12}x_6 - \omega_{13}(x_3+2\sqrt{3}x_8)-\frac{\alpha_3}{6}x_5(1+4\sqrt{3}x_8)\nonumber\\
    \dot{x}_6 &= -\omega_{13}x_2-\omega_{12}x_5-\frac{\alpha_3}{6}x_6(1+4\sqrt{3}x_8)\nonumber\\
    \dot{x}_7 &= -\omega_{13}x_1+\omega_{12}x_4+\omega_{23}(x_3-2\sqrt{3}x_8)-\frac{\alpha_3}{6}x_7(1+4\sqrt{3}x_8)\nonumber\\
    \dot{x}_8&=\frac{\sqrt{3}}{2}\bigg[\omega_{13}x_5+\omega_{23}x_7+\frac{2}{9}\alpha_3(1-\sqrt{3}x_8(1+2\sqrt{3}x_8))\bigg]
\end{alignat}
The functional $\mathcal{F}$ is given by $\mathcal{F}=-\frac{\alpha_3}{3}x_8(1-2\sqrt{3}x_8)$. The dynamical Hamiltonian is given by
\begin{alignat}{1}\label{eq:ham}
    \mathcal{H}&=\sum_{i=1}^8 p_i\dot{x_i}+\mathcal{F}.
\end{alignat}
The canonically conjugate momenta can be derived by Hamilton's equations
\begin{equation}
    p_i=-\frac{\partial\mathcal{H}}{\partial x_i}.
\end{equation}
Thus we obtain the coupled equations:
\begin{alignat}{1}
    \dot{p}_1 &= -\frac{\alpha_3}{3}(1-2\sqrt{3}x_8)p_1+\omega_{23}p_5+\omega_{13}p_7\nonumber\\
    \dot{p}_2 &= -\frac{\alpha_3}{3}(1-2\sqrt{3}x_8)p_2-2\omega_{12}p_3-\omega_{23}p_4 +\omega_{13}p_6\nonumber\\
    \dot{p}_3 &= 2\omega_{12}p_2-\frac{\alpha_3}{3}(1-2\sqrt{3}x_8)p_3+\omega_{13}p_5-\omega_{23}p_7\nonumber\\
    \dot{p}_4 &= \omega_{23}p_2+\frac{\alpha_3}{6} (1+4\sqrt{3}x_8)p_4\nonumber\\
    \dot{p}_5 &= \omega_{23} p_1 -\omega_{13}p_3+\frac{\alpha_3}{6} (1+4\sqrt{3}x_8)p_5+\omega_{12}p_6-\frac{\sqrt{3}}{2}\omega_{13}p_8\nonumber\\
    \dot{p}_6&=-\omega_{13}p_2-\omega_{12}p_5+\frac{\alpha_3}{6}(1+4\sqrt{3}x_8)p_6\nonumber\\
    \dot{p}_7 &= -\omega_{13}p_1+\omega_{23}p_3+\omega_{12}p_4+\frac{\alpha_3}{6}(1+4\sqrt{3}x_8)p_7-\frac{\sqrt{3}}{2}\omega_{23}p_7\nonumber\\
    \dot{p}_8 &=\frac{2}{\sqrt{3}}\alpha_3(x_1p_1+x_2p_2+x_3p_3+x_4p_4+x_5p_5+x_6p_6+x_7p_7+2x_8p_8)\nonumber\\
    &+2\sqrt{3}(\omega_{13}p_5+\omega_{23}p_7)+\frac{\alpha_3}{3}(p_8+1)-\frac{4}{\sqrt{3}}\alpha_3x_8.
\end{alignat}
The dynamics of the position coordinates of the qutrit with time are shown in Fig. \ref{fig:single_dynamics}. When the detection frequency is less compared to all the transition frequencies of the system, the dynamics shows continuous oscillations, Fig. \ref{fig:single_dynamics} (a). In an intermediate frequency, the system shows oscillations for some time, after which, it gets arrested in a particular state, Fig. \ref{fig:single_dynamics} (b). When the detection frequency is higher compared to all the transition frequencies of the system, the Zeno regime sets in, Fig. \ref{fig:single_dynamics} (c). Each coordinate freezes at a particular value around a time $t=6$ and the system does not evolve any further. 

\begin{figure*}[ht!]
    \centering
    \subfloat[]{\includegraphics[width=0.37\textwidth]{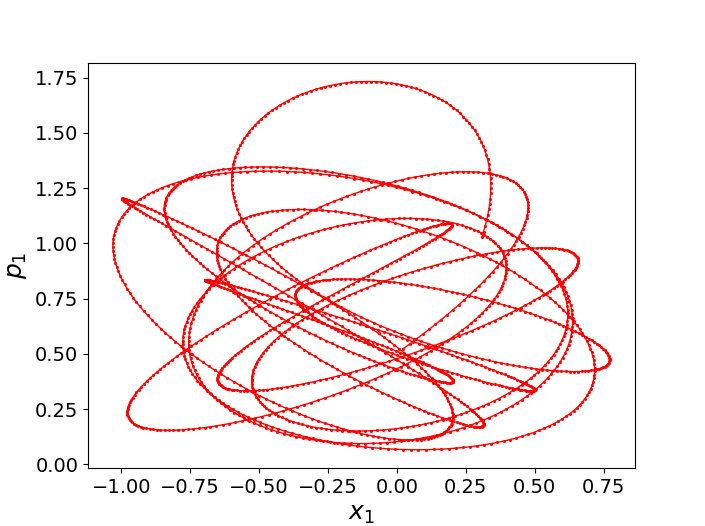}} 
    \subfloat[]{\includegraphics[width=0.35\textwidth]{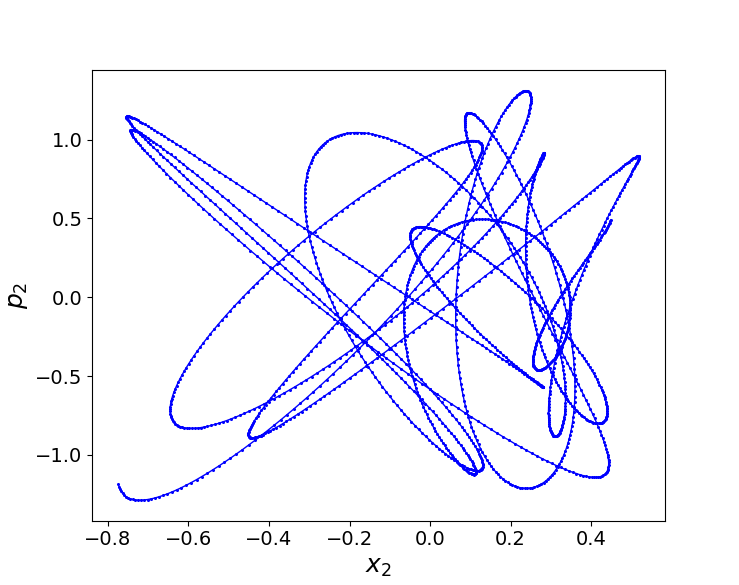}}
    \subfloat[]{\includegraphics[width=0.35\textwidth]{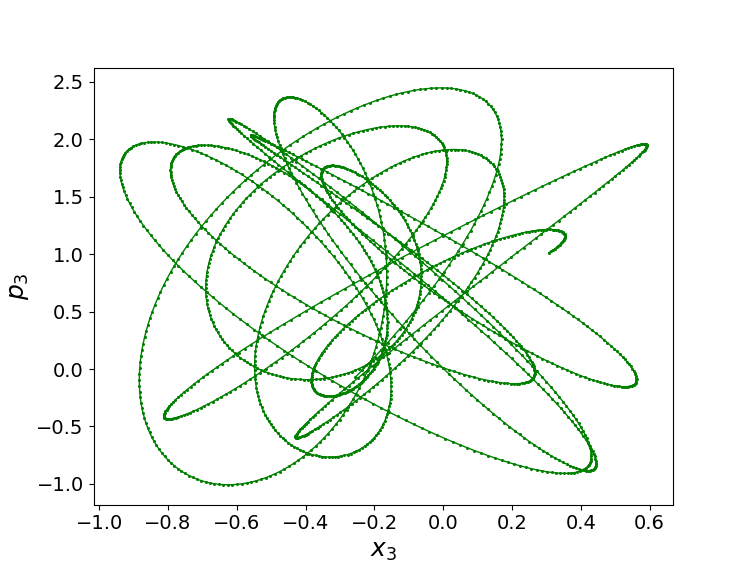}}\\
    \subfloat[]{\includegraphics[width=0.35\textwidth]{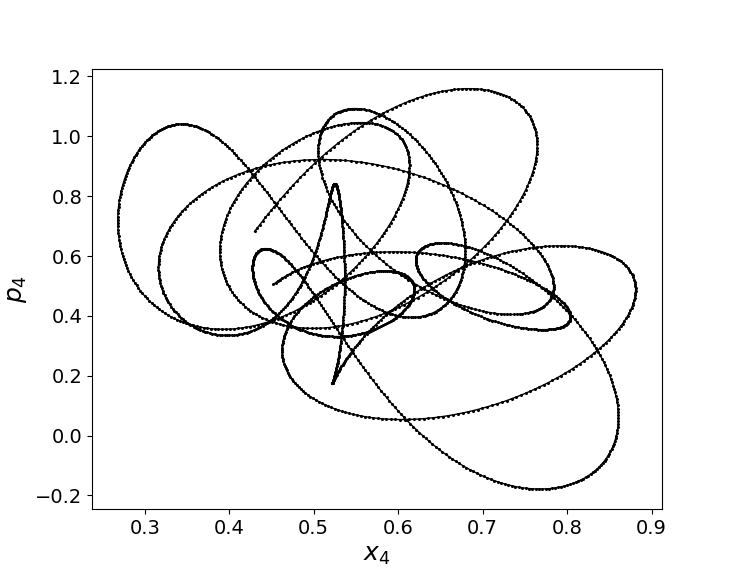}}    
    \subfloat[]{\includegraphics[width=0.35\textwidth]{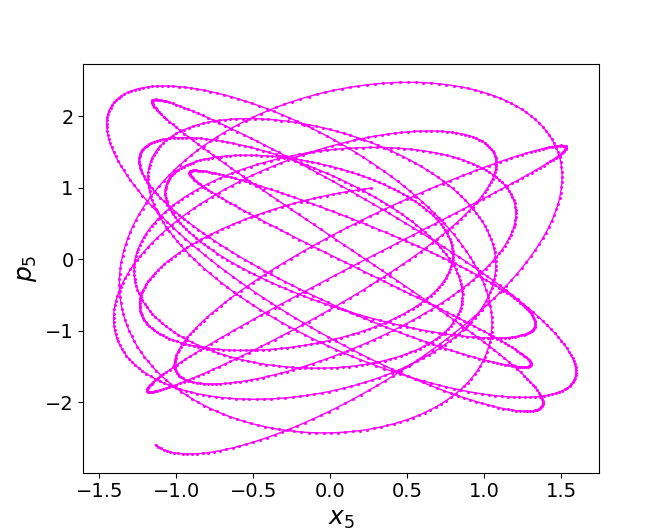}}
    \subfloat[]{\includegraphics[width=0.35\textwidth]{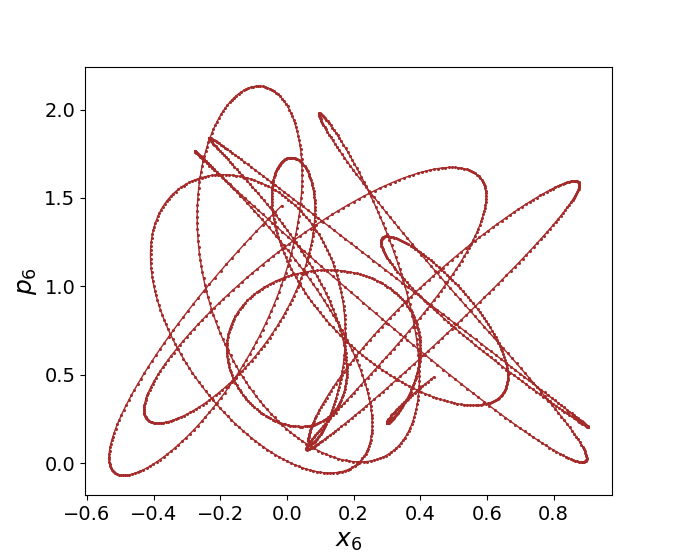}}\\
    \subfloat[]{\includegraphics[width=0.4\textwidth]{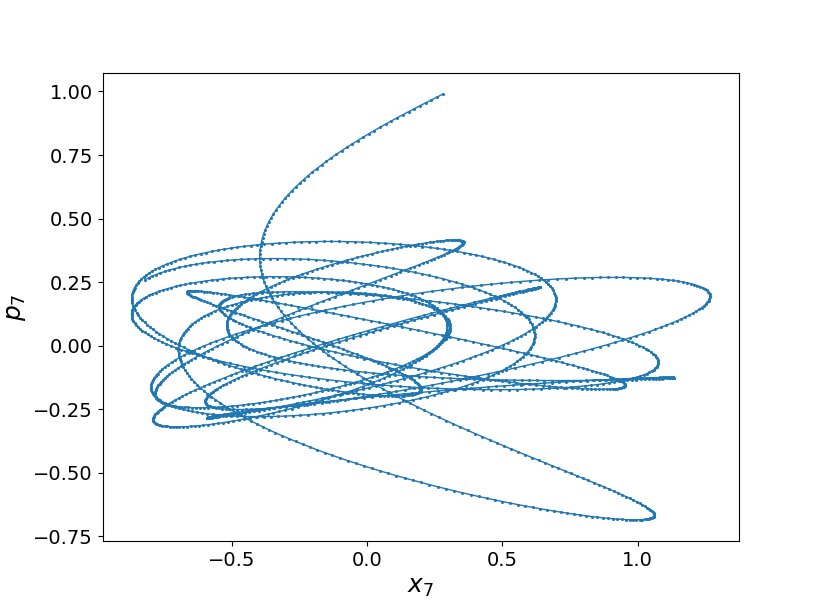}}
    \subfloat[]{\includegraphics[width=0.38\textwidth]{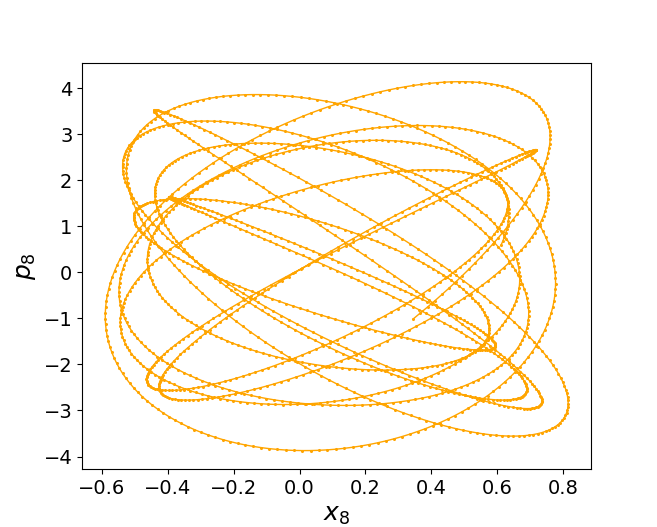}}
    \caption{For $\alpha_3=0.1$, the phase space dynamics of variable (a) $x_1$, (b) $x_2$, (c) $x_3$, (d) $x_4$, (e) $x_5$, (f) $x_6$, (g) $x_7$ and (h) $x_8$ is shown for chosen initial conditions $x_1=x_3=x_5=x_7=0.3$, $x_2=x_4=x_6=0.4$, $p_1=p_3=p_5=p_7=1$ and $p_2=p_4=p_6=p_8=0.5$ for a total time of $t=20$. Figure shows evolution of the system in phase space in the non-Zeno regime.}
\label{fig:x_p-single_al_0.1}
\end{figure*}

\begin{figure*}[ht!]
    \centering
    \subfloat[]{\includegraphics[width=0.37\textwidth]{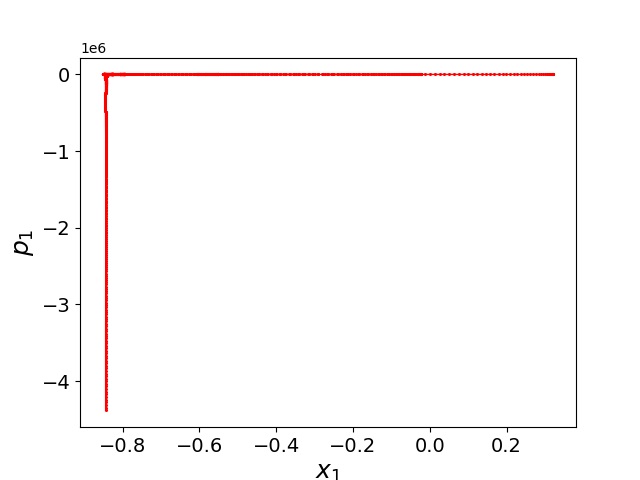}} 
    \subfloat[]{\includegraphics[width=0.35\textwidth]{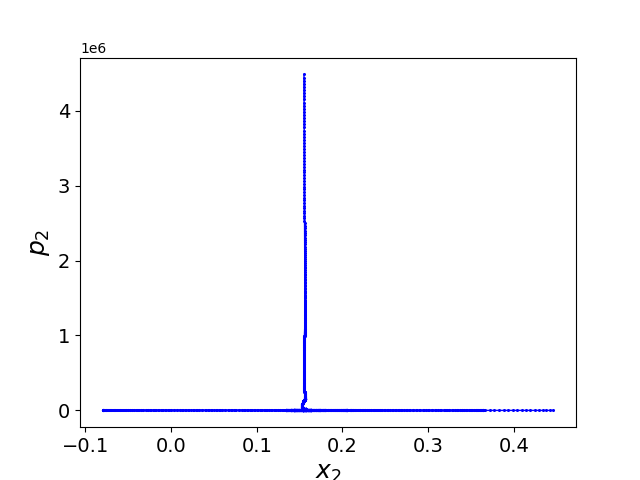}}
    \subfloat[]{\includegraphics[width=0.35\textwidth]{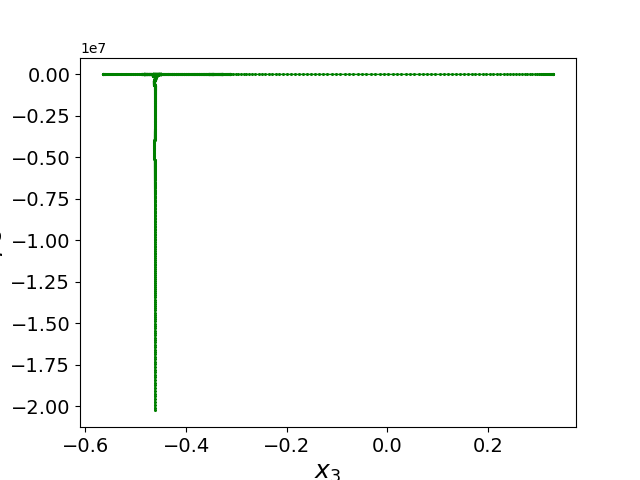}}\\
    \subfloat[]{\includegraphics[width=0.35\textwidth]{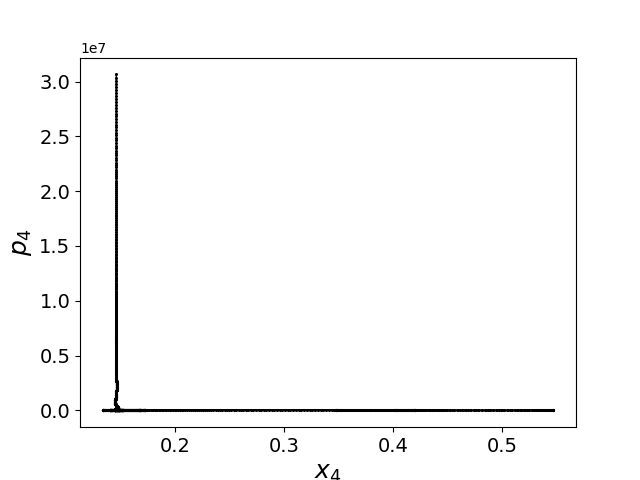}}    
    \subfloat[]{\includegraphics[width=0.35\textwidth]{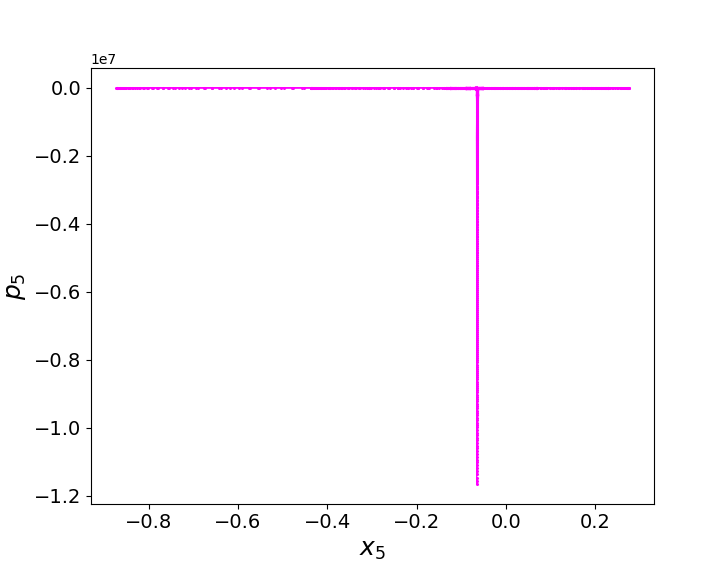}}
    \subfloat[]{\includegraphics[width=0.35\textwidth]{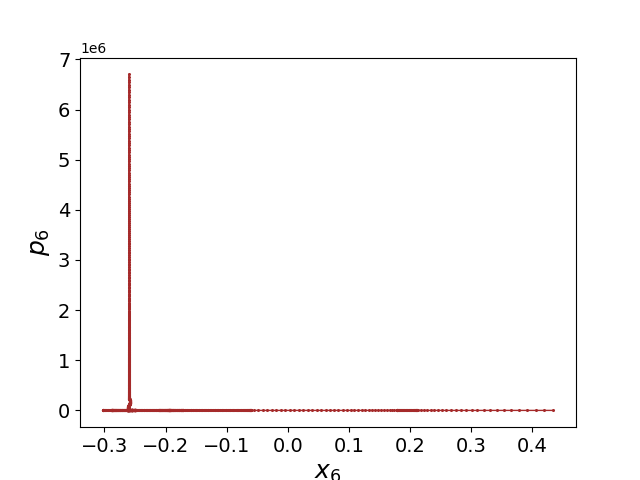}}\\
    \subfloat[]{\includegraphics[width=0.4\textwidth]{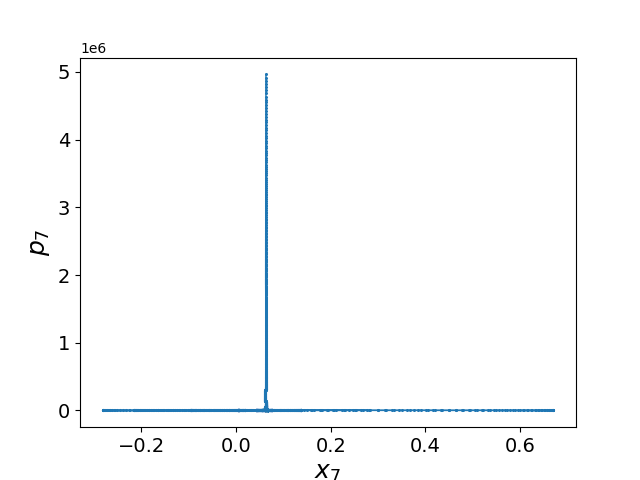}}
    \subfloat[]{\includegraphics[width=0.38\textwidth]{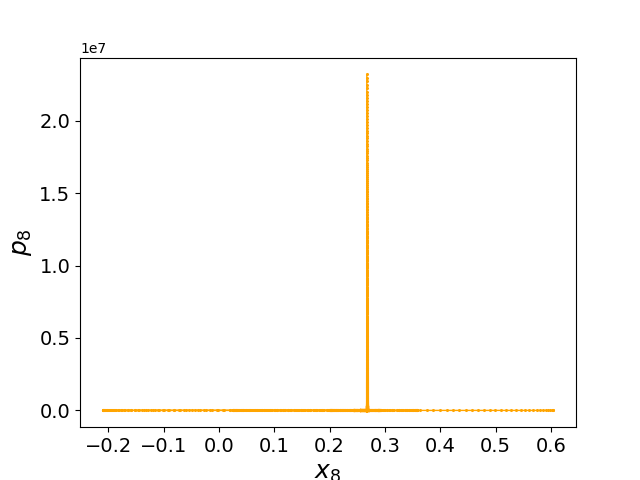}}
    \caption{For $\alpha_3=1.7$, the phase space dynamics of variable (a) $x_1$, (b) $x_2$, (c) $x_3$, (d) $x_4$, (e) $x_5$, (f) $x_6$, (g) $x_7$ and (h) $x_8$ is shown for the same initial conditions as in Fig. \ref{fig:x_p-single_al_0.1} for a total time of $t=20$. The figure shows that when the Zeno regime has completely set in, i.e., the detector frequency $\alpha_3>\omega_{12},\omega_{23}, \omega_{13}$, the dynamics is completely arrested at a particular point. Since the system is delocalised in the momentum coordinates and localised in the position coordinates, there is a saddle point. The qutrit, in the Zeno regime gets shelved at the critical point.}
\label{fig:x_p-single_al_1.7}
\end{figure*}

The phase space dynamics of the qutrit are plotted in Figs. \ref{fig:x_p-single_al_0.1} and \ref{fig:x_p-single_al_1.7}, for a frequency lower and higher than the transition frequencies, respectively. In Fig. \ref{fig:x_p-single_al_0.1}, for each coordinate, the qutrit shows evolution in the phase-space. However, in the Zeno regime, Fig. \ref{fig:x_p-single_al_1.7}, it is evident that localization in $x(p)$ is accompanied by delocalization of $p(x)$. This shows that the system is shelved to a state. In terms of stability, localization in $x$ or $p$ corresponds to stability along that coordinate. It is clear that both $x$ and $p$ are not stable simultaneously, hence the points are saddle points, as in \cite{krjj}.  

\subsection{Creating a cNOT gate}
The three-level system can be used as a control and the ancilla as a target such that when the system is in $\ket{1}$ or $\ket{2}$, it does nothing to the ancilla (ancilla stays in initial state $\ket{0}_{(n)}$, whereas flips the ancilla to $\ket{1}_{(n)}$ when qutrit is in $\ket{3}$. Such a gate can be represented as
\begin{alignat}{1}
    \text{cNOT}&=(\ket{1}\bra{1}+\ket{2}\bra{2})\otimes \hat{\mathbb{I}} + \ket{3}\bra{3}\otimes \sigma_x^{(n)}.
\end{alignat}
The states on which the cNOT acts are $\ket{1,0}$, $\ket{2,0}$ or $\ket{3,0}$, where the first state is the qutrit state which controls the target ancilla initially in the state $\ket{0}$. When cNOT acts on $\ket{3,0}$, it gives $\ket{3,1}$ and leaves the others unchanged. 

\subsection{Dense coding and teleportation}

Some of the applications of entangled pairs are dense coding and teleportation. Dense coding uses one quantum bit together with a shared EPR pair to encode and transmit two classical bits \cite{rieffel}. Without using entanglement, only one classical bit of information can be extracted. Teleportaion is the opposite of dense coding as it uses two classical bits to transmit the state of an unknown qubit. The initial setup for both includes two parties, Alice and Bob who wish to communicate. Each is sent one of the entangled particles of an EPR pair
\begin{equation}\label{eq:in_state}
    \ket{\psi_0}=\frac{1}{\sqrt{2}} (\ket{0}_A\ket{0}_B+\ket{1}_A\ket{1}_B).
\end{equation}
Each can perform transformations only on their particle unless they send over their particle. 
$\\ \\$
\noindent
\textit{Dense coding}: Alice wants to transmit the state of two classical bits encoding one of the numbers $\{0,1,2,3\}$, depending on which, she performs one of the transformations $\{I,X,Y,Z\}$ on her qubit of $\ket{\psi_0}$. The resulting state is shown in table \ref{tab:dc1_qubit}.
\begin{table}[tbph!]
    \centering
    \begin{tabular}{c|c|c}
          Value & Transformation & New state\\
         \hline
      0 & ($I\otimes I$)$\ket{\psi_0}$ & $\frac{1}{\sqrt{2}}(\ket{00}+\ket{11})$\\
      1 & ($X\otimes I$)$\ket{\psi_0}$ & $\frac{1}{\sqrt{2}}(\ket{10}+\ket{01})$\\
      2 & ($Z\otimes I$)$\ket{\psi_0}$ & $\frac{1}{\sqrt{2}}(\ket{00}-\ket{11})$\\
      3 & ($Y\otimes I$)$\ket{\psi_0}$ & $\frac{1}{\sqrt{2}}(-\ket{10}+\ket{01})$\\
    \end{tabular}
    \caption{The value of the classical bit is encoded by Alice and sent to Bob.}
    \label{tab:dc1_qubit}
\end{table}
Bob decodes the information in two steps: cNOT to the entangled pair followed by Hadamard $H$ on the first qubit:
\begin{table}[tbph!]
    \centering
    \begin{tabular}{c|c|c}
    State & cNOT & $H\otimes I$\\
        \hline
        $\frac{1}{\sqrt{2}}(\ket{00}+\ket{11})$ & $\frac{1}{\sqrt{2}}(\ket{00}+\ket{10})$ = $\frac{1}{\sqrt{2}}(\ket{0}+\ket{1})\otimes\ket{0}$ & $\ket{00}$ \\
        $\frac{1}{\sqrt{2}}(\ket{10}+\ket{01})$ & $\frac{1}{\sqrt{2}}(\ket{11}+\ket{01})$ = $\frac{1}{\sqrt{2}}(\ket{1}+\ket{0})\otimes\ket{1}$ & $\ket{01}$\\
        $\frac{1}{\sqrt{2}}(\ket{00}-\ket{11})$ & $\frac{1}{\sqrt{2}}(\ket{00}-\ket{10})$ = $\frac{1}{\sqrt{2}}(\ket{0}-\ket{1})\otimes\ket{0}$ & $\ket{10}$ \\
        $\frac{1}{\sqrt{2}}(-\ket{10}+\ket{01})$ & $\frac{1}{\sqrt{2}}(-\ket{11}+\ket{01})$ = $\frac{1}{\sqrt{2}}(-\ket{1}+\ket{0})\otimes\ket{1}$ & $\ket{11}$\\
    \end{tabular}
    \caption{cNOT followed by Hadamard is applied by Bob with Alice's qubit as control.}
    \label{tab:dc2_qubit}
\end{table}
Bob finally measures the two qubits to obtain the binary encoding sent by Alice.
$\\ \\$
\noindent
\textit{Quantum teleportation}: Due to the no-cloning theorem, the original state is destroyed and finally created at the target, hence the name teleportation. Alice has an qubit with unknown state $\ket{\phi}=a\ket{0}+b\ket{1}$. Both Alice and Bob share a part of the EPR pair just like in dense coding \eqref{eq:in_state}. The initial state is then the three-qubit state:
\begin{alignat}{1}\label{eq:in_state_tele}
  \ket{\psi}\otimes\ket{\psi_0}&=\frac{1}{\sqrt{2}}(a\ket{0}\otimes(\ket{00}+\ket{11})+b\ket{1}\otimes(\ket{00}+\ket{11}))\nonumber\\
  &=\frac{1}{\sqrt{2}}(a\ket{000}+a\ket{011}+b\ket{100}+b\ket{111}).
\end{alignat}
Alice controls the first two qubits and Bob controls the third. Alice uses the decoding step used by Bob in dense coding to the first two qubits in \eqref{eq:in_state_tele}, i.e., cNOT on first two followed by Hadamard on first qubit
\begin{alignat}{1}
    (H\otimes I\otimes I)&(\text{cNOT}\otimes I)(\ket{\psi}\otimes\ket{\psi})\nonumber\\
    &=(H\otimes I\otimes I)\frac{1}{\sqrt{2}}(a\ket{000}+a\ket{011}+b\ket{110}+b\ket{101})\nonumber\\
    &=\frac{1}{2}[a(\ket{000}+\ket{011}+\ket{100}+\ket{111})+b(\ket{010}+\ket{001}-\ket{110}-\ket{101})]\nonumber\\
    &=\frac{1}{2}(\ket{00}(a\ket{0}+b\ket{1})+\ket{01}(a\ket{1}+b\ket{0})+\ket{10}(a\ket{0}-b\ket{1})+\ket{11}(a\ket{1}-b\ket{0})).
\end{alignat}
Upon measuring the first two qubits, Alice obtains one of the four states $\ket{00}$, $\ket{01}$, $\ket{10}$ or $\ket{11}$, depending upon which, Bob's qubit is projected to one of the four states $a\ket{0}+b\ket{1}$, $a\ket{1}+b\ket{0}$, $a\ket{0}-b\ket{1}$ or $a\ket{1}-b\ket{0}$. Alice sends her result as two classical bits to Bob. The original state $\ket{\phi}$ is contained in Bob's qubits. Upon receiving the two bits, Bob reconstructs the state by applying decoding transformation to his qubit:
\begin{table}[htbp!]
    \centering
    \begin{tabular}{c|c|c}
    State & Bits received & Decoding\\
    \hline
       $a\ket{0}+b\ket{1}$  & 00 & $I$\\
       $a\ket{1}+b\ket{0}$  & 01 & $X$\\
       $a\ket{0}-b\ket{1}$  & 10 & $Z$\\
        $a\ket{1}-b\ket{0}$ & 11 & $Y$\\
    \end{tabular}
    \caption{Decoding the state by Bob using the bits received from Alice.}
    \label{tab:dc3_qubit}
\end{table}
Bob will finally have the qubit Alice wished to send.

\subsection{Applications of entanglement using three-level system}

We have considered a three-level system where the third level is being monitored by an ancilla. For communication and teleportation using the qutrit, we need to have two of the states acting as ground and the third, which is being monitored as the higher level. This will enable us to create a cNOT gate for the qutrit. Further, we need the regular Pauli operators corresponding to this setup, such that the bit-flip operator acts on the states as
\begin{align*}
    X_{13}\ket{1}=\ket{3}, \quad X_{23}\ket{2}=\ket{3}, \quad X_{13+23}\ket{3}= \frac{\ket{1}+\ket{2}}{\sqrt{2}}.
\end{align*}
Hence, the operators may be written as
\begin{align*}
    X_{13}=\begin{bmatrix}
        0 & 0 & 1\\
        0 & 0 & 0\\
        1 & 0 & 0\\
    \end{bmatrix} \quad X_{23}=\begin{bmatrix}
        0 & 0 & 0\\
        0 & 0 & 1\\
        0 & 1 & 0\\
    \end{bmatrix}.
\end{align*}
The resulting $X$ operator read as
\begin{align*}
    X=\frac{X_{13}+X_{23}}{\sqrt{2}}=\frac{1}{\sqrt{2}}\begin{bmatrix}
        0 & 0 & 1\\
        0 & 0 & 1\\
        1 & 1 & 0\\
    \end{bmatrix}.
\end{align*}
We have
\begin{align*}
    X\ket{1}=\frac{1}{\sqrt{2}}\begin{bmatrix}
        0\\
        0\\
        1\\
    \end{bmatrix}=\frac{1}{\sqrt{2}}\ket{3}, \quad X\ket{2}=\frac{1}{\sqrt{2}}\begin{bmatrix}
        0\\
        0\\
        1\\
    \end{bmatrix}=\frac{1}{\sqrt{2}}\ket{3} \quad \text{and} \quad X\ket{3}=\frac{\ket{1}+\ket{2}}{\sqrt{2}}.
\end{align*}
Similarly, the $Y$ operator is
\begin{align*}
    Y=\frac{1}{\sqrt{2}}\begin{bmatrix}
        0 & 0 & 1\\
        0 & 0 & 1\\
        -1 & -1 & 0\\
    \end{bmatrix},
\end{align*}
with
\begin{align*}
    Y\ket{1}=\frac{1}{\sqrt{2}}\begin{bmatrix}
        0\\
        0\\
        -1\\
    \end{bmatrix}=-\frac{1}{\sqrt{2}}\ket{3}, \quad Y\ket{2}=\frac{1}{\sqrt{2}}\begin{bmatrix}
        0\\
        0\\
        -1\\
    \end{bmatrix}=-\frac{1}{\sqrt{2}}\ket{3} \quad \text{and} \quad Y\ket{3}=\frac{\ket{1}+\ket{2}}{\sqrt{2}}.
\end{align*}
The phase operator should act as
\begin{align*}
    Z\left(\frac{\ket{1}+\ket{2}}{\sqrt{2}}\right)=\left(\frac{\ket{1}+\ket{2}}{\sqrt{2}}\right)\quad \text{and}\quad Z\ket{3}=-\ket{3}.
\end{align*}
That is,
\begin{align*}
    Z=\frac{1}{\sqrt{2}}\begin{bmatrix}
        1 & 0 & 0\\
        0 & 1 & 0\\
        0 & 0 & -\sqrt{2}\\
    \end{bmatrix}.
\end{align*}
The cNOT gate is given by
\begin{align*}
    \text{cNOT}=(\ket{1}\bra{1}+\ket{2}\bra{2})\otimes I^{(n)}+\ket{3}\bra{3}\otimes \sigma_{x}^{(n)},
\end{align*}
where superscript $(n)$ represents the ancilla. To find the Hadamard operator, note that
\begin{align*}
    \ket{\psi_0}&=\frac{1}{\sqrt{2}}\left(\frac{(\ket{1}+\ket{2})}{\sqrt{2}}+\ket{3}\right)\nonumber\\
    H&\frac{1}{\sqrt{2}}\left(\frac{(\ket{1}+\ket{2})}{\sqrt{2}}+\ket{3}\right)=\frac{\ket{1}+\ket{2}}{\sqrt{2}}\nonumber\\
    H&\frac{1}{\sqrt{2}}\left(\frac{(\ket{1}+\ket{2})}{\sqrt{2}}-\ket{3}\right)=\ket{3}.
\end{align*}
These are effected by the Hadamard gate:
\begin{align*}
    H=\frac{1}{2\sqrt{2}}\begin{bmatrix}
        1 & 1 & \sqrt{2}\\
        1 & 1 & \sqrt{2}\\
        \sqrt{2} & \sqrt{2} & -2\\
    \end{bmatrix}.
\end{align*}
Now we have a set of operators at our disposal, acting as gates on this three-level system for dense coding and teleportation.$\\ \\$
\noindent
\textit{Dense coding}: Alice encodes the digits $\{0,1,2,3\}$ in state $\ket{\psi_0}$ and performs transformations on her part of the state. Let the states of ancilla be $\{\ket{g},\ket{e}\}$, the eigenstates of $\sigma_z$. These are entangled with the qutrit to parallel the EPR pair of qubits.
\begin{table}[tbph!]
    \centering
    \begin{tabular}{c|c|c}
          Value & Transformation & New state\\
         \hline
      0 & ($I\otimes I$)$\ket{\psi_0}$ & $\frac{1}{\sqrt{2}}\left(\frac{(\ket{1}+\ket{2})}{\sqrt{2}}\ket{g}+\ket{3}\ket{e}\right)$\\
      1 & ($X\otimes I$)$\ket{\psi_0}$ & $\frac{1}{\sqrt{2}}\left(\ket{3}\ket{g}+\frac{(\ket{1}+\ket{2})}{\sqrt{2}}\ket{e}\right)$\\
      2 & ($Z\otimes I$)$\ket{\psi_0}$ & $\frac{1}{\sqrt{2}}\left(\frac{(\ket{1}+\ket{2})}{\sqrt{2}}\ket{g}-\ket{3}\ket{e}\right)$\\
      3 & ($Y\otimes I$)$\ket{\psi_0}$ & $\frac{1}{\sqrt{2}}\left(-\ket{3}\ket{g}+\frac{(\ket{1}+\ket{2})}{\sqrt{2}}\ket{e}\right)$\\
    \end{tabular}
    \caption{The value of the classical bit is encoded by Alice and sent to Bob.}
    \label{tab:dc1_qutrit}
\end{table}
Then, Bob decodes using cNOT followed by Hadamard on the (first) qutrit. Here, the cNOT has control as the three-level system and target as a two level system. Hence, the flip operator will be the usual 2D Pauli $\sigma_x$. This is shown in table \ref{tab:dc2_qutrit}.$\\$
\begin{table}[tbph!]
    \centering
    \begin{tabular}{c|c|c}
    State & cNOT & $H\otimes I$\\
        \hline
        $\frac{1}{\sqrt{2}}\left(\frac{(\ket{1}+\ket{2})}{\sqrt{2}}\ket{g}+\ket{3}\ket{e}\right)$ & $\frac{1}{\sqrt{2}}\left(\frac{(\ket{1}+\ket{2})}{\sqrt{2}}+\ket{3}\right)\ket{g}$ & $\left(\frac{(\ket{1}+\ket{2})}{\sqrt{2}}\right)\ket{g}$ \\
        $\frac{1}{\sqrt{2}}\left(\ket{3}\ket{g}+\frac{(\ket{1}+\ket{2})}{\sqrt{2}}\ket{e}\right)$ & $\frac{1}{\sqrt{2}}\left(\ket{3}+\frac{(\ket{1}+\ket{2})}{\sqrt{2}}\right)\ket{e}$ & $\left(\frac{(\ket{1}+\ket{2})}{\sqrt{2}}\right)\ket{e}$\\
        $\frac{1}{\sqrt{2}}\left(\frac{(\ket{1}+\ket{2})}{\sqrt{2}}\ket{g}-\ket{3}\ket{e}\right)$ & $\frac{1}{\sqrt{2}}\left(\frac{(\ket{1}+\ket{2})}{\sqrt{2}}-\ket{3}\right)\ket{g}$ & $\ket{3}\ket{g}$ \\
        $\frac{1}{\sqrt{2}}\left(-\ket{3}\ket{g}+\frac{(\ket{1}+\ket{2})}{\sqrt{2}}\ket{e}\right)$ & $\frac{1}{\sqrt{2}}\left(-\ket{3}+\frac{(\ket{1}+\ket{2})}{\sqrt{2}}\right)\ket{e}$ & $\ket{3}\ket{e}$\\
    \end{tabular}
    \caption{cNOT followed by Hadamard is applied by Bob with Alice's qutrit as control.}
    \label{tab:dc2_qutrit}
\end{table}

\noindent
\textit{Teleportation}: Alice has an unknown qubit $\ket{\phi}=a\ket{g}+b\ket{e}$ (ancilla). She wants to send this to Bob through a classical channel. They each share a part of the state
\begin{equation}
    \ket{\psi_0}=\frac{1}{\sqrt{2}}\left[\frac{\ket{11}+\ket{12}+\ket{21}+\ket{22}}{2}+\ket{33}\right],
\end{equation}
so that the combined state initially is
\begin{alignat}{1}\label{eq:in_tele}
    \ket{\phi}\otimes\ket{\psi_0}&=\frac{1}{2\sqrt{2}}[a(\ket{g11}+\ket{g12}+\ket{g21}+\ket{g22}+2\ket{g33})\nonumber\\
    &+b(\ket{e11}+\ket{e12}+\ket{e21}+\ket{e22}+2\ket{e33})].
\end{alignat}
Alice controls the first two states in the tensor product in \eqref{eq:in_tele} and Bob controls the third state. For the decoding step, Alice applies cNOT ($\ket{g}\bra{g}\otimes I_3+\ket{e}\bra{e}\otimes X_3$) on the first two states of the product followed by Hadamard on the first
\begin{alignat}{1}
    (H_2\otimes I\otimes I)&(\text{cNOT}\otimes I)(\ket{\phi}\otimes\ket{\psi_0})\nonumber\\
    =&(H_2\otimes I\otimes I)\frac{1}{2\sqrt{2}}[a(\ket{g11}+\ket{g12}+\ket{g21}+\ket{g22}+2\ket{g33})\nonumber\\
    &+\sqrt{2}b(\ket{e31}+\ket{e32}+\frac{1}{\sqrt{2}}(\ket{e13}+\ket{e23}))]\nonumber\\
    =&\frac{1}{4}[a(\ket{g11}+\ket{e11}+\ket{g12}+\ket{e12}+\ket{g21}+\ket{e21}+\ket{g22}+\ket{e22}+2\ket{g33}+2\ket{e33})\nonumber\\
    &+\sqrt{2}b(\ket{g31}-\ket{e31}+\ket{g32}-\ket{e32}+\ket{g13}-\ket{e13}+\ket{g23}-\ket{e23})]\nonumber\\
    =&\frac{1}{2\sqrt{2}}\bigg[\ket{g1}\left(a\frac{(\ket{1}+\ket{2})}{\sqrt{2}}+b\ket{3}\right)+\ket{e1}\left(a\frac{(\ket{1}+\ket{2})}{\sqrt{2}}-b\ket{3}\right)\nonumber\\
    &+\ket{g2}(a\frac{(\ket{1}+\ket{2})}{\sqrt{2}}+b\ket{3})+\ket{e2}\left(a\frac{(\ket{1}+\ket{2})}{\sqrt{2}}-b\ket{3}\right)\nonumber\\
    &+\ket{g3}\left(\sqrt{2}a\ket{3}+ \sqrt{2}b\frac{(\ket{1}+\ket{2})}{\sqrt{2}}\right)+\ket{e3}\left(\sqrt{2}a\ket{3}- \sqrt{2}b\frac{(\ket{1}+\ket{2})}{\sqrt{2}}\right)\bigg].
\end{alignat}
Thus the final encoded state is
\begin{alignat}{1}
    \ket{\psi}_f&=\frac{1}{2}\bigg[\ket{g}\left(\frac{\ket{1}+\ket{2}}{\sqrt{2}}\right)\bigg\{a\left(\frac{\ket{1}+\ket{2}}{\sqrt{2}}\right)+b\ket{3}\bigg\}+\ket{e}\left(\frac{\ket{1}+\ket{2}}{\sqrt{2}}\right)\bigg\{a\left(\frac{\ket{1}+\ket{2}}{\sqrt{2}}\right)-b\ket{3}\bigg\}\nonumber\\
    &+\ket{g}\ket{3}\bigg\{a\ket{3}+b\left(\frac{\ket{1}+\ket{2}}{\sqrt{2}}\right)\bigg\}+\ket{e}\ket{3}\bigg\{a\ket{3}-b\left(\frac{\ket{1}+\ket{2}}{\sqrt{2}}\right)\bigg\}\bigg]
\end{alignat}
Upon measuring the first two states, Alice will obtain one of the four states mentioned in the first column of Tab. \ref{tab:tele_3}, which she sends as two classical bits to Bob. Upon receiving them, Bob reconstructs the state by applying a decoding transformation (\ref{tab:tele_3}) to his part of the product state which contains the unknown state $\ket{\phi}$. Thus Bob will finally have the qubit state Alice wanted to send.
\begin{table}[htbp!]
    \centering
    \begin{tabular}{c|c|c}
    State & Bits received & Decoding\\
    \hline
       $\frac{1}{\sqrt{2}}\left(a\left(\frac{\ket{1}+\ket{2}}{\sqrt{2}}\right)+b\ket{3}\right)$ & $g,12$ & $I$\\
       $\frac{1}{\sqrt{2}}\left(a\ket{3}+b\left(\frac{\ket{1}+\ket{2}}{\sqrt{2}}\right)\right)$  & $g,3$ & $X$\\
       $\frac{1}{\sqrt{2}}\left(a\left(\frac{\ket{1}+\ket{2}}{\sqrt{2}}\right)-b\ket{3}\right)$  & $e,12$ & $Z$\\
        $\frac{1}{\sqrt{2}}\left(a\ket{3}-b\left(\frac{\ket{1}+\ket{2}}{\sqrt{2}}\right)\right)$ & $e,3$ & $Y$\\
    \end{tabular}
    \caption{Decoding the state by Bob using the bits received from Alice.}
    \label{tab:tele_3}
\end{table}

\subsection{Monitoring two levels}
\begin{figure*}[ht!]
    \centering
    \subfloat[]{\includegraphics[width=0.6\textwidth]{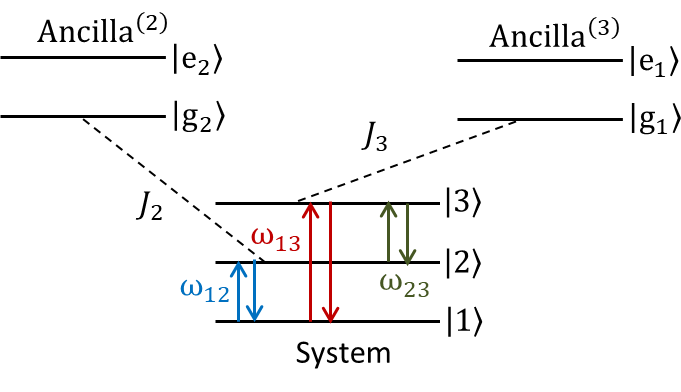}} 
    \caption{A qutrit is interacting with two ancillae or detectors. Ancilla$^{(2)}$ (Ancilla$^{(3)}$) is a two-level system which monitors the state $\ket{2}$ ($\ket{3}$) of the qutrit with a coupling strength $J_2$ ($J_3$). The transition frequencies of the three-levels are $\omega_{12}$ between states $\ket{1}$ and $\ket{2}$, $\omega_{23}$ between states $\ket{2}$ and $\ket{3}$ and $\omega_{13}$ between states $\ket{1}$ and $\ket{3}$.}
\label{fig:system-two-an}
\end{figure*}

\begin{figure*}[ht!]
    \centering
    \subfloat[]{\includegraphics[width=0.35\textwidth]{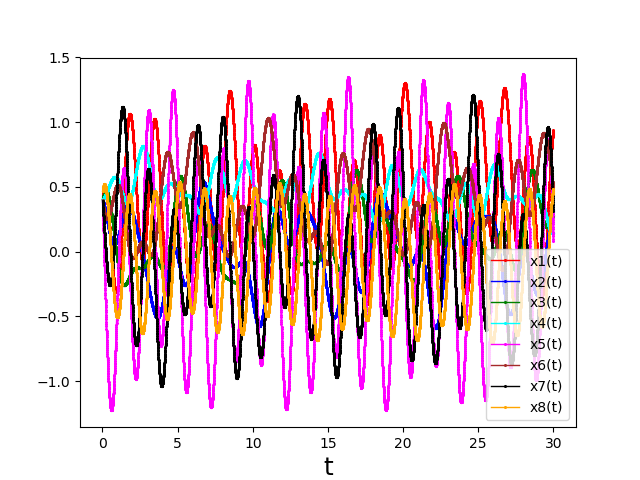}} 
    \subfloat[]{\includegraphics[width=0.35\textwidth]{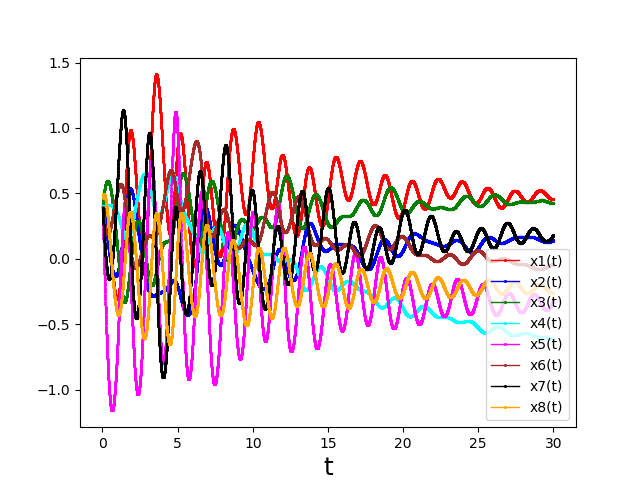}} 
    \subfloat[]{\includegraphics[width=0.35\textwidth]{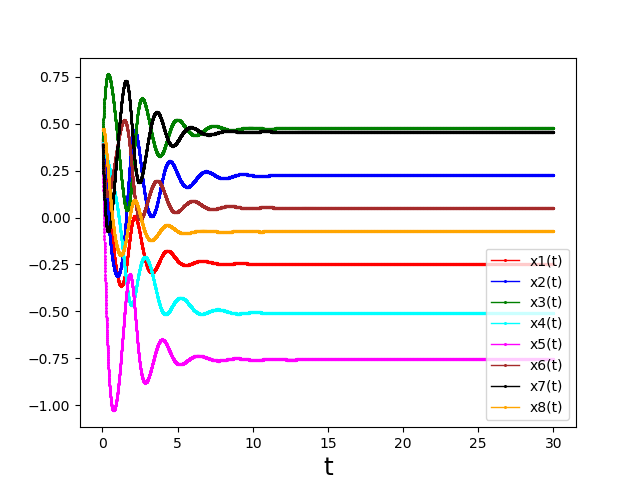}}
    \caption{Figure shows the dynamics of the three-level system, the variation of its 8 variables plotted with time when the third level is being monitored. The initial conditions have been chosen as $x_1=x_3=x_5=x_7=0.3$, $x_2=x_4=x_6=0.5$ and $x_8=\sqrt{(4/3)^2-(\sum_i x_i)^2}$, where $i=1,2,\dots,7$. The Rabi frequencies of the three levels is chosen to be $\omega_{12}=0.6$, $\omega_{23}=1$ and $\omega_{13}=1.6$. The system is being monitored in three frequency ranges, (a) $\alpha2=\alpha_3=0.1$, (b) $\alpha2=\alpha_3=0.7$ and (c) $\alpha2=\alpha_3=1.7$. In fig. (a), there are usual coherent oscillation. In (b), the system begins to freeze fairly early at a particular state. In fig. (c) the Zeno regime has set in.}
\label{fig:al2al3dynamics}
\end{figure*}

Consider a qutrit interacting with two ancillae. The ancillae are again two-level systems, one of which monitor the state $\ket{2}$ whereas the other monitors the state $\ket{3}$ as shown in Fig. \ref{fig:system-two-an}. The interaction strength between qutrit and ancilla monitoring $\ket{2}$ ($\ket{3}$) is $J_2=\sqrt{\frac{\alpha_2}{\delta t}}$ ($J_3=\sqrt{\frac{\alpha_3}{\delta t}}$).
The Hamiltonian for this system can be given as
\begin{alignat}{1}
    H&=\omega_{12}(\ket{1}\bra{2}+\ket{2}\bra{1})+\omega_{23}(\ket{2}\bra{3}+\ket{3}\bra{2})+\omega_{13}(\ket{1}\bra{3}+\ket{3}\bra{1})+ H_{s-d},
\end{alignat}
where
\begin{alignat}{1}
    H_{s-d} = J_2\ket{2}\bra{2}\otimes\sigma_y^{(2)}\otimes\mathbb{I}^{(3)} + J_3 \ket{3}\bra{3} \otimes \mathbb{I}^{(2)}\otimes \sigma_y^{(3)} + (J_2 \ket{2}\bra{2}+ J_3 \ket{3}\bra{3}) \otimes \sigma_y^{(2)} \otimes \sigma_y^{(3)}.
\end{alignat}
The Kraus operators are given by
\begin{alignat}{1}\label{eq:kraus_2lev}
    \mathcal{M}_r&=\bra{r_1 r_2}\exp{[-\iota H_{s-d} \delta t]}\ket{00}\nonumber\\
    \mathcal{M}_{00} &= \mathbb{I} -J_2^2\ket{2}\bra{2} (\delta t)^2 -J_3^2\ket{3}\bra{3} (\delta t)^2\nonumber\\
    \mathcal{M}_{01} &= -J_3\ket{3}\bra{3} \delta t -\iota J_2^2 \ket{2}\bra{2} (\delta t)^2\nonumber\\
    \mathcal{M}_{10} &= -J_2\ket{2}\bra{2} \delta t -\iota J_3^2 \ket{3}\bra{3} (\delta t)^2\nonumber\\
    \mathcal{M}_{11} &= \iota (J_2 \ket{2}\bra{2} +J_3 \ket{3}\bra{3}) \delta t.
\end{alignat}
So we have a $2\times 2$ Kraus operator matrix. The unitary evolution of qutrit under system Hamiltonian $H_s$ and measurement postselected on $r=00$, we obtain 8 coupled dynamic equations from the density matrix
\begin{equation}
    \rho(t+\delta t)=\frac{\mathcal{M}_{00}\mathcal{U}\rho \mathcal{U}^\dagger\mathcal{M}_{00}^\dagger}{Tr[\mathcal{M}_{00} \mathcal{U}\rho \mathcal{U}^\dagger\mathcal{M}_{00}^\dagger]}.
\end{equation}
These equations are
\begin{alignat}{1}
    \dot{x}_1 &= -\alpha_2 x_1x_3 +\omega_{23}x_5 +\omega_{13}x_7+\frac{1}{3}(\alpha_2-2\alpha_3)x_1(2\sqrt{3}x_8-1)\nonumber\\
    \dot{x}_2 &= - \left[2\omega_{12}x_3 + \alpha_2 x_2x_3 + \omega_{23} x_4-\omega_{13} x_6 -\frac{1}{3}(\alpha_2-2\alpha_3)x_2(2\sqrt{3}x_8-1) \right]\nonumber\\
    \dot{x}_3 &= \frac{1}{3} [6\omega_{12}x_2 + 2\alpha_3 x_3 +3 \omega_{23} x_5 -3\omega_{23}x_7-4\sqrt{3}\alpha_3 x_3 x_8-\alpha_2(1+x_3)(-2+3x_3-2\sqrt{3}x_8)]\nonumber\\
    \dot{x}_4 &=\frac{1}{3}\left[3\omega_{23}x_2-3\omega_{12}x_7 +\alpha_2 x_4 (2-3x_3+2\sqrt{3}x_8) -\alpha_3 x_4 (1+4\sqrt{3} x_8) \right] \nonumber\\
    \dot{x}_5 &=\frac{1}{3}[-3\omega_{23}x_1+(2\alpha_2-\alpha_3-3\alpha_2 x_3)x_5 +3\omega_{12}x_6+2\sqrt{3}(\alpha_2-2\alpha_3)x_5x_8-3\omega_{13}(x_3+2\sqrt{3}x_8)]\nonumber\\
    \dot{x}_6 &= \left[-\omega_{13} x_2 - \omega_{12}x_5 - \frac{1}{3}x_6 (\alpha_2+\alpha_3+3\alpha_2 x_3 - 2\sqrt{3}\alpha_2 x_8 +4\sqrt{3}\alpha_3 x_8) \right]\nonumber\\
    \dot{x}_7 &= \left[-\omega_{13}x_1 +\omega_{12}x_4 -\frac{1}{3} (\alpha_2+\alpha_3 +3 \alpha_2 x_3)x_7 +\frac{2}{\sqrt{3}}(\alpha_2-2\alpha_3)x_7x_8+\omega_{23} (x_3-2\sqrt{3}x_8) \right]\nonumber\\
    \dot{x}_8 &= \frac{1}{6\sqrt{3}} [4 \alpha_3 + 9 \omega _{13} x_5+9 \omega_{23}x_7-4\alpha_3x_8(\sqrt{3}+6x_8)+\alpha_3(-2+3x_3+2\sqrt{3}(1-3x_3)x_8+12x_8^2)]
\end{alignat}
The functional incorporating the backaction is $\mathcal{F}=\alpha_2x_3-\frac{2}{3}(\alpha_2+\alpha_3+\sqrt{3}\alpha_2x_8-2\sqrt{3}\alpha_3x_8)$. The corresponding conjugate momenta are
\begin{alignat}{1}
    p_1&=\alpha_2x_3p_1-\frac{1}{3}(\alpha_2-2\alpha_3)(2\sqrt{3}x_8-1)p_1+\omega_{23}p_5+\omega_{13}p_7\nonumber\\
    p_2&=\alpha_2x_3 p_2-\frac{1}{3}(\alpha_2-2\alpha_3)(2\sqrt{3}x_8-1)p_2-2\omega_{12}p_3-\omega_{23}p_4+\omega_{13}p_6\nonumber\\
    p_3&=\alpha_2x_1p_1+2\omega_{12}p_2+\alpha_2x_2p_2-\frac{2}{3}\alpha_3p_3+\frac{4\sqrt{3}}{3}\alpha_3x_8p_3+\frac{\alpha_2}{3}(1+6x_3-2\sqrt{3}x_8)p_3\nonumber\\
    &+\alpha_2x_4p_4+\alpha_2x_5p_5+\omega_{13}p_5+\alpha_2x_6p_6+\alpha_2x_7p_7-\omega_{23}p_7-\frac{\alpha_3}{2\sqrt{3}}p_8+x_8p_8-\alpha_2\nonumber\\
    p_4&=\omega_{23}p_2-\frac{\alpha_2}{3}(2-3x_3+2\sqrt{3}x_8)p_4+\frac{\alpha_3}{3}(1+4\sqrt{3}x_8)p_4-\omega_{12}p_7\nonumber\\
    p_5&=-\omega_{23}p_1-\omega_{23}p_3-\frac{1}{3}(2\alpha_2-\alpha_3-3\alpha_2x_3)p_5-\frac{2\sqrt{3}}{3}(\alpha_2-2\alpha_3)x_8p_5+\omega_{12}p_6-\frac{\sqrt{3}}{2}\omega_{13}p_8\nonumber\\
    p_6&=-\omega_{13}p_2-\omega_{12}p_5+\frac{1}{3}(\alpha_2+\alpha_3+3\alpha_2x_3-2\sqrt{3}\alpha_2x_8+4\sqrt{3}\alpha_3x_8)p_6\nonumber\\
    p_7&=-\omega_{13}p_1+\omega_{23}p_3+\omega_{12}p_4+\frac{1}{3}(\alpha_2+\alpha_3+3\alpha_2x_3)p_7-\frac{2}{\sqrt{3}}(\alpha_2-2\alpha_3)x_8p_7-\frac{\sqrt{3}}{2}\omega_{23}p_8\nonumber\\
    p_8&=-\frac{2}{\sqrt{3}}(\alpha_2-2\alpha_3)(x_1p_1+x_2p_2+x_3p_3+x_4p_4+x_5p_5+x_6p_6+x_7p_7-1\textcolor{red}{?})-\frac{2}{\sqrt{3}}\alpha_2p_3\nonumber\\
    &+2\sqrt{3}(\omega_{13}p_5+\omega_{23}p_7)+\frac{2}{3}\alpha_3(1+2\sqrt{3}x_8)p_8-\frac{\alpha_3}{3}(1-3x_3)p_8-\frac{4}{\sqrt{3}}\alpha_3x_8p_8.
\end{alignat}
The dynamics of the position coordinates of the qutrit with time are shown in Fig. \ref{fig:al2al3dynamics}. When the detection frequencies of the two detectors are less compared to all the transition frequencies of the system, the dynamics shows continuous oscillations, Fig. \ref{fig:al2al3dynamics} (a). In an intermediate frequency range, the system shows oscillations for some time, after which, it gets arrested in a particular state, Fig. \ref{fig:al2al3dynamics} (b). When the detection frequency is higher compared to all the transition frequencies of the system, the Zeno regime sets in, Fig. \ref{fig:al2al3dynamics} (c). Each coordinate freezes at a particular value around a time $t=6$, just as in the previous section where a single state was being monitored. 

\begin{figure*}[htbp!]
    \centering
    \subfloat[]{\includegraphics[width=0.36\textwidth]{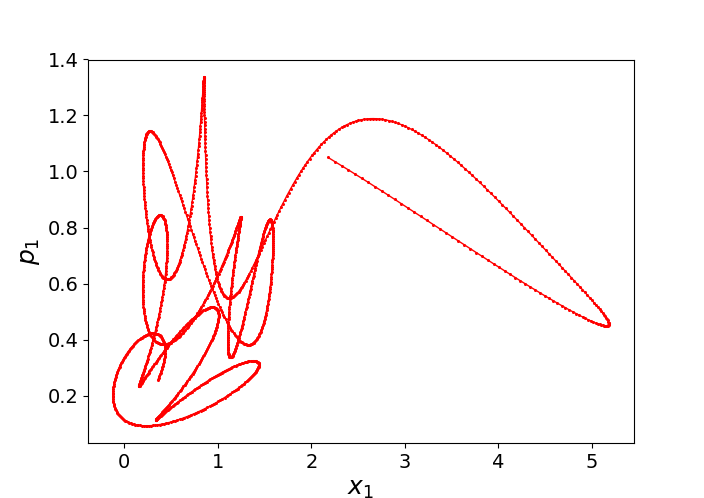}} 
    \subfloat[]{\includegraphics[width=0.33\textwidth]{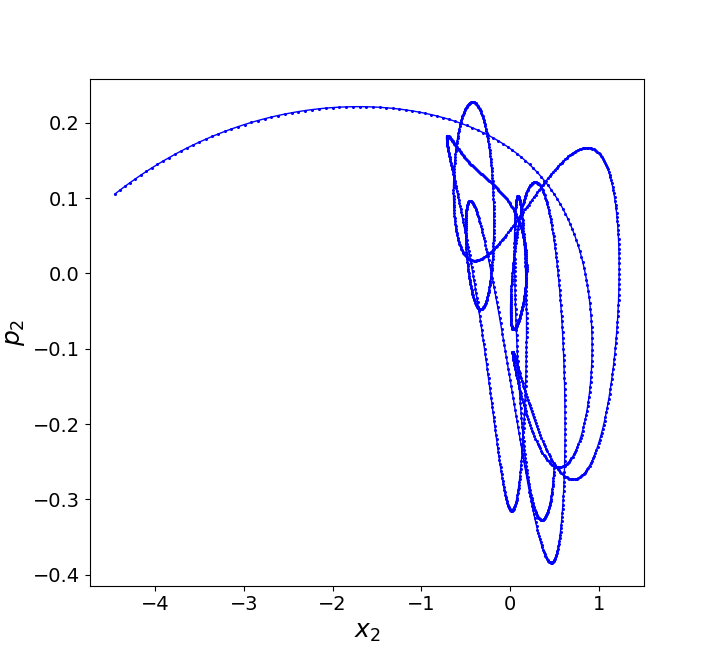}}
    \subfloat[]{\includegraphics[width=0.35\textwidth]{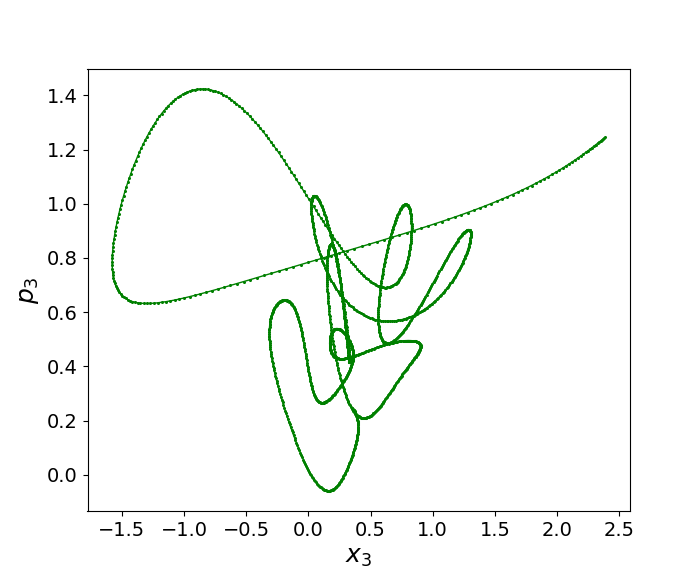}}\\
    \subfloat[]{\includegraphics[width=0.35\textwidth]{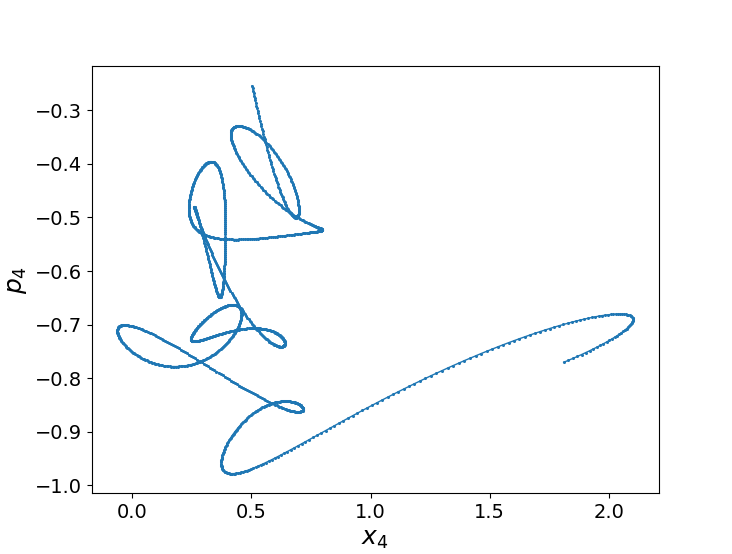}}    
    \subfloat[]{\includegraphics[width=0.35\textwidth]{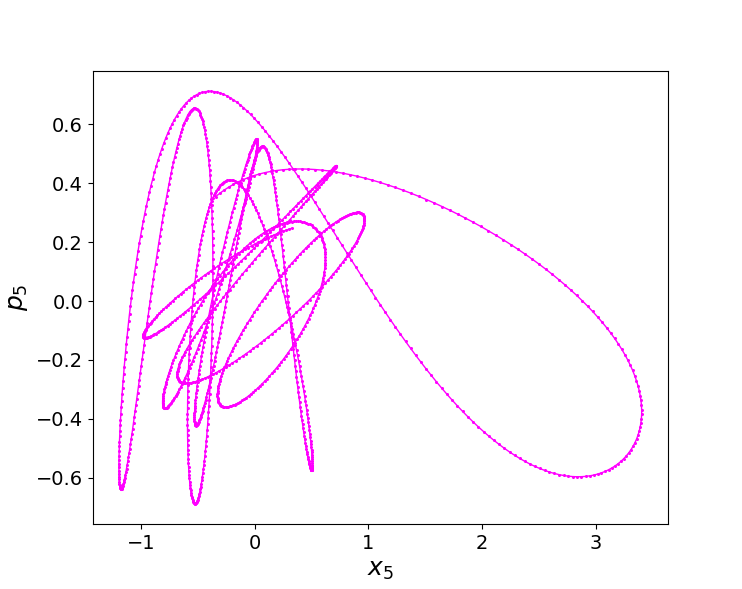}}
    \subfloat[]{\includegraphics[width=0.35\textwidth]{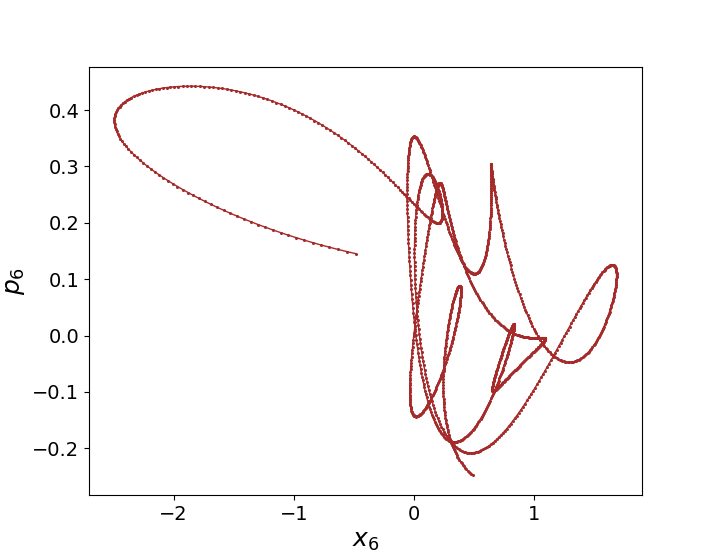}}\\
    \subfloat[]{\includegraphics[width=0.4\textwidth]{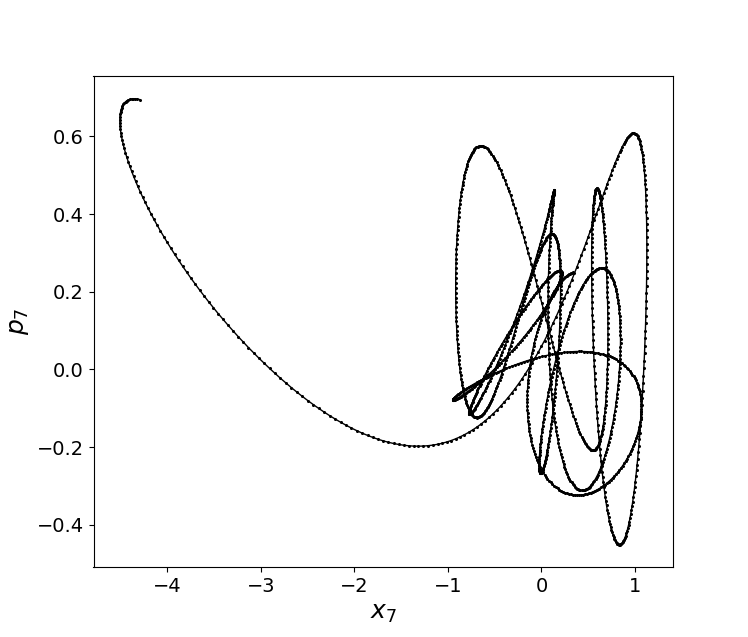}}
    \subfloat[]{\includegraphics[width=0.4\textwidth]{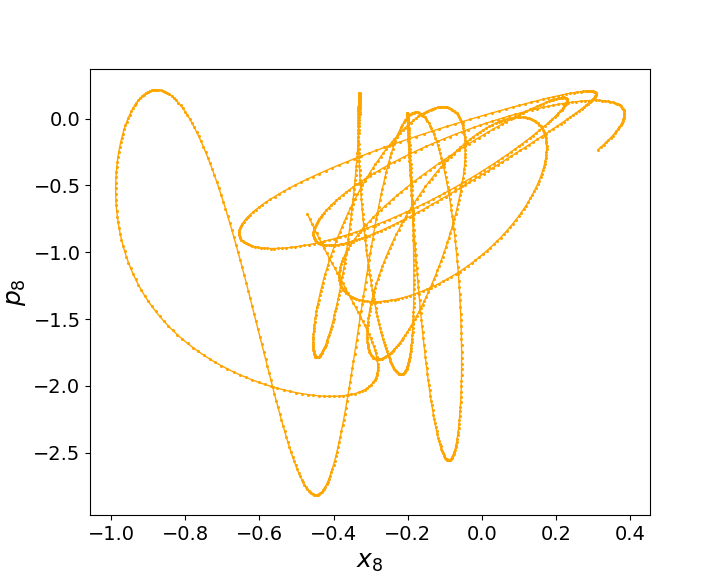}}
    \caption{For $\alpha_2=0.1$ and $\alpha_3=0.2$, the phase space dynamics of variable (a) $x_1$, (b) $x_2$, (c) $x_3$, (d) $x_4$, (e) $x_5$, (f) $x_6$, (g) $x_7$ and (h) $x_8$ is shown for chosen initial conditions $x_1=x_3=x_5=x_7=0.35$, $x_2=x_4=x_6=0.5$, $p_1=p_3=p_5=p_7=0.25$ and $p_2=p_4=p_6=p_8=-0.25$ for a total time of $t=15$. The figure shows that when the Zeno regime has completely set in, i.e., the detector frequency $\alpha_3>\omega_{12},\omega_{23}, \omega_{13}$, the dynamics is completely arrested and follows uncertainty principle. There is a saddle point at which the qutrit gets shelved in the position coordinates and the hence are deloclaised in the momentum coordinates.}
\label{fig:x_p_al_1.9_1.7}
\end{figure*}

\begin{figure*}[htbp!]
    \centering
    \subfloat[]{\includegraphics[width=0.4\textwidth]{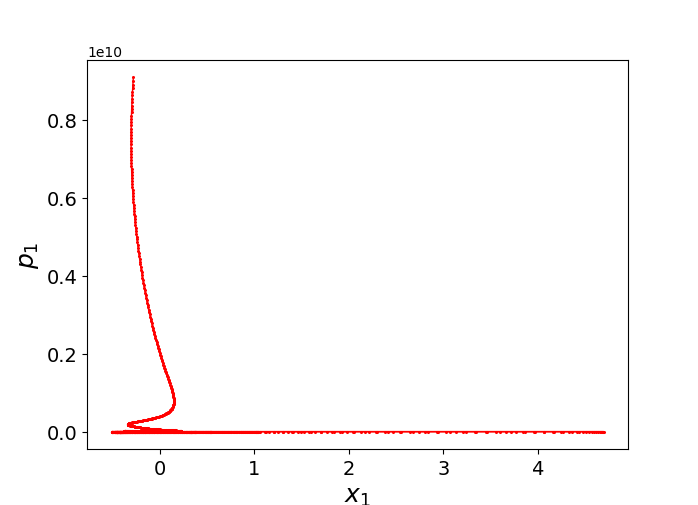}} 
    \subfloat[]{\includegraphics[width=0.36\textwidth]{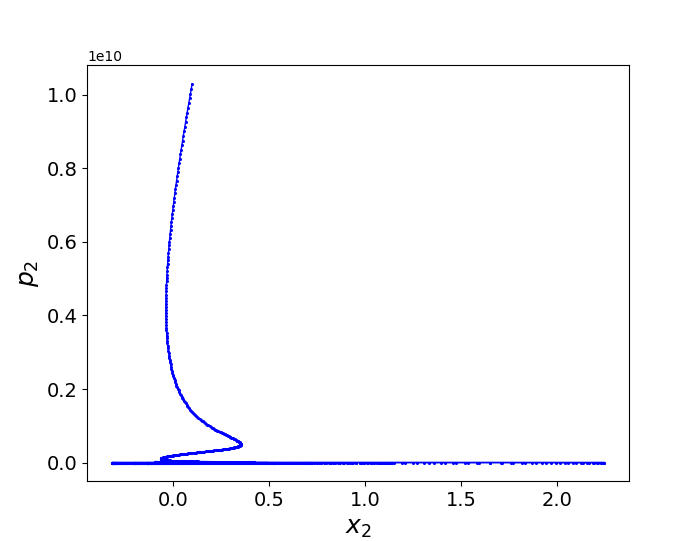}}
    \subfloat[]{\includegraphics[width=0.35\textwidth]{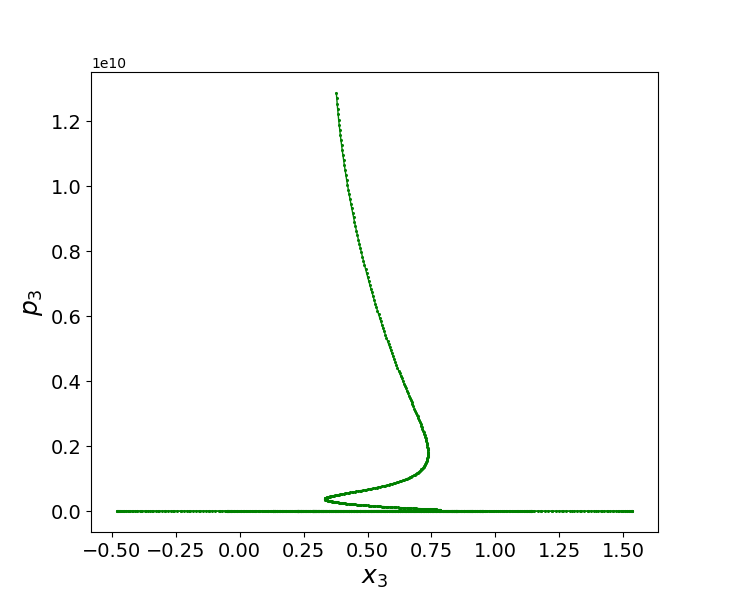}}\\
    \subfloat[]{\includegraphics[width=0.35\textwidth]{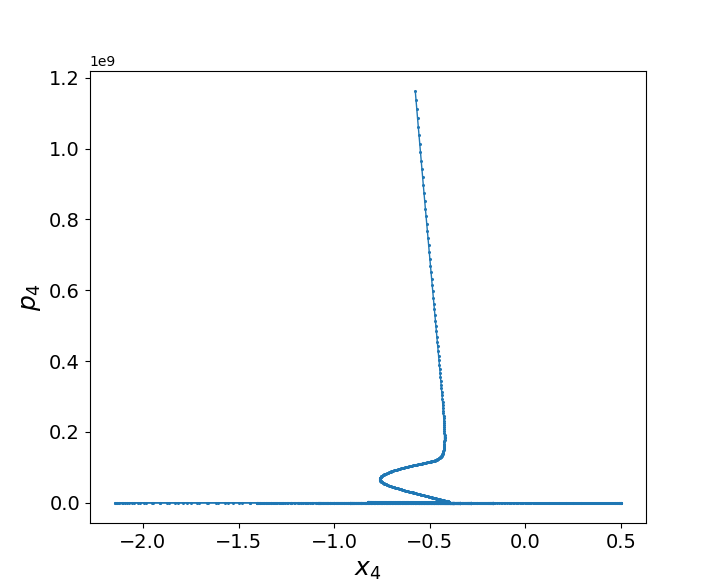}}    
    \subfloat[]{\includegraphics[width=0.35\textwidth]{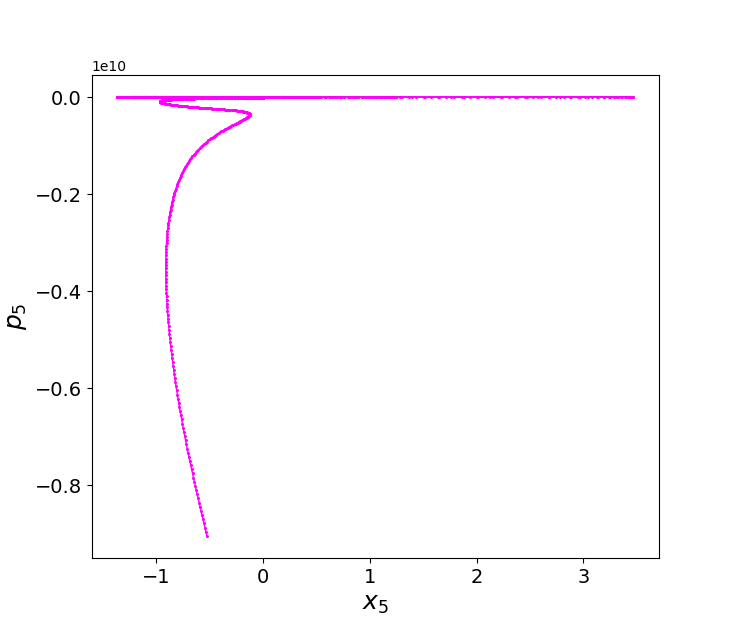}}
    \subfloat[]{\includegraphics[width=0.35\textwidth]{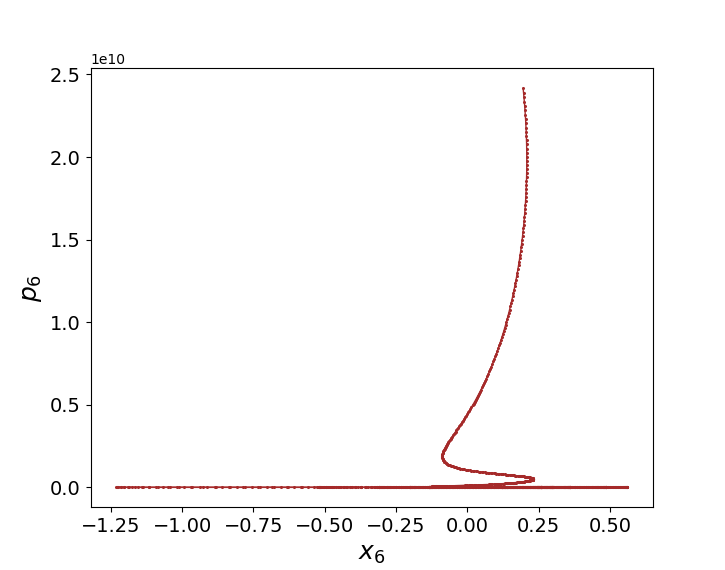}}\\
    \subfloat[]{\includegraphics[width=0.4\textwidth]{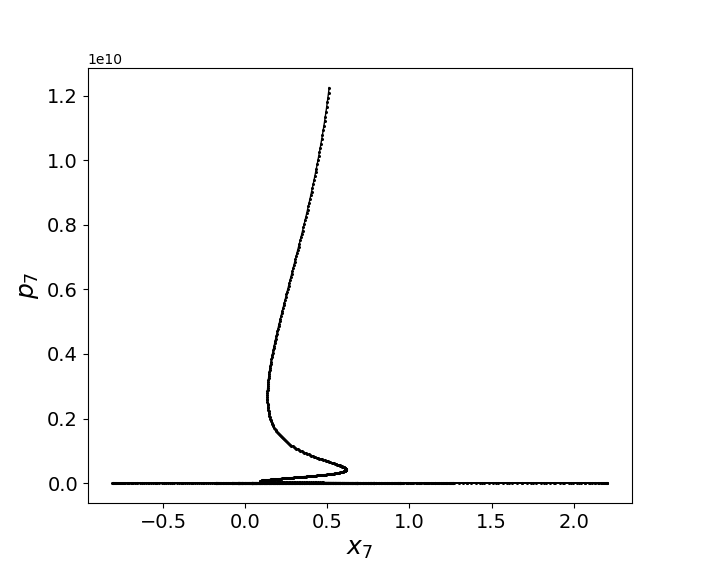}}
    \subfloat[]{\includegraphics[width=0.4\textwidth]{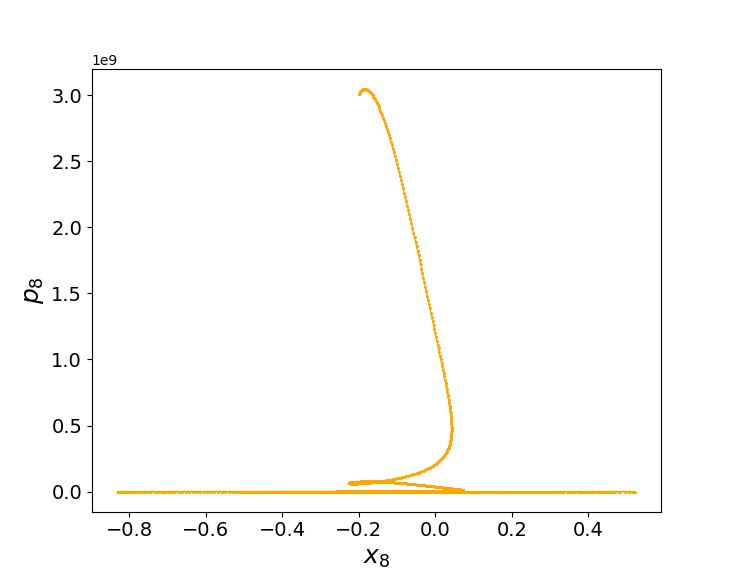}}
    \caption{For $\alpha_2=1.9$ and $\alpha_3=1.7$, the phase space dynamics of variable (a) $x_1$, (b) $x_2$, (c) $x_3$, (d) $x_4$, (e) $x_5$, (f) $x_6$, (g) $x_7$ and (h) $x_8$ is shown for chosen initial conditions $x_1=x_3=x_5=x_7=0.35$, $x_2=x_4=x_6=0.5$, $p_1=p_3=p_5=p_7=0.25$ and $p_2=p_4=p_6=p_8=-0.25$ for a total time of $t=15$. Figure shows evolution of the system in phase space in the non-Zeno regime.}
\label{fig:x_p_al_0.1_0.2}
\end{figure*}

The phase space dynamics of the qutrit are plotted in Figs. \ref{fig:x_p_al_0.1_0.2} and \ref{fig:x_p_al_1.9_1.7}, for frequencies of both the detectors lower and higher than the transition frequencies, respectively. In Fig. \ref{fig:x_p_al_0.1_0.2}, for each coordinate, the qutrit shows evolution in the phase-space. However, in the Zeno regime, Fig. \ref{fig:x_p_al_1.9_1.7}, the system follows uncertainty principle - as soon as the position coordinates are fixed at a particular value, the uncertainty in the momentum coordinates peaks. This also shows that there is a saddle point. The qutrit gets shelved in the position coordinates and the is delocalised in the momentum coordinates.
\section{Creating a Toffoli gate}
The Kraus operators in \eqref{eq:kraus_2lev} indicate that the system may be in state $1$ ($M_{00}$), state 2 ($M_{10}$), state 3 ($M_{01}$) or in a combination of 2 and 3, i.e., anywhere but not in state 1 ($M_{23}$). This can be interpreted as an operator
\begin{alignat}{1}
    T &= \ket{1}\bra{1}\otimes \ket{1}\bra{1}\otimes (\mathbb{I}\otimes\mathbb{I}) + \ket{2}\bra{2}\otimes \ket{1}\bra{1}\otimes (X\otimes\mathbb{I}) \nonumber \\ &+\ket{1}\bra{1}\otimes\ket{3}\bra{3}\otimes (\mathbb{I}\otimes X) + \ket{2}\bra{2} \otimes\ket{3}\bra{3}\otimes (X\otimes X).
\end{alignat}
Consider $\mathbb{I}\otimes\mathbb{I}$, $\mathbb{I}\otimes X$ and $x\otimes \mathbb{I}$ as giving an outcome of 0 and $x\otimes X$ equivalent to producing an outcome of 1. The setup can then be interpreted as a Toffoli gate. For instance, if control is $(1,1)$ and target is 0, the state is $\ket{1,1,(00)\equiv 0}$. If control is $(2,3)$ and target is 1, the state is $\ket{2,3,(11)\equiv 1}$.

\section{Concluding remarks}
Control of qutrit is shown by monitoring one or two levels. Due to the Quantum Zeno Effect, the state of the system is shown to shelve to a state other than the states of the three-level system. Treatment to a three-level system takes us out of Pauli algebra, here we have Gell-Mann matrices. In addition, we write a new set of operators to realise the cNOT gate with the qutrit as the control and the two-level ancilla as the target. With these operators, the applications of entanglement have been realised in a three-level system in dense coding and teleportation for the purpose of quantum communication. Application of the system to universal gates allows us to manipulate the states. In general, for $N-$level system also, the conclusion will hold good. 
\vskip 0.5 truecm
\noindent
{\bf Data Availability Statement}: No Data associated in the manuscript 
\vskip 0.25 truecm
\noindent
{\bf Conflict of interests}: Authors declare no conflict of interest.
\newpage
\appendixpage
\section*{Appendix: Density matrix of \texorpdfstring{$N$}{Lg}-level system}\label{app:AppI}
An $N-$level system is defined by a Bloch vector whose components are expectation values of some observables \cite{kimura}. The number of observables needed to identify the state are $N^2-1$. These correspond to $N^2-1$ independent parameters used to define a Hermitian density matrix operator $\hat{\rho}$ with a constraint, $\text{Tr} \hat{\rho}=1$. Choosing the generators of $SU(N)$ for the observables $\hat{x}_i$, the density matrix is determined from their expectation values $\langle\hat{x}_i\rangle$'s as
\begin{equation}
    \rho=\frac{1}{N}\hat{\mathbb{I}}_N + \frac{1}{2} \sum_{i=1}^{N^2-1} \langle\hat{x}_i\rangle\hat{x}_i.
\end{equation}
The properties of the density matrix associated with a Hilbert space $\mathcal{H}_N$ is given as
\begin{align*}\label{eq:rho_prop}
    \rho\in\mathcal{L}(\mathcal{H}_N) : (\rm{i}) \rm{Tr} \rho=1 \quad(\rm{ii}) \rho = \rho^\dagger \quad (\rm{iii}) \rho_i \geq 0,
\end{align*}
where $\mathcal{L}$ is the space of linear operators on $\mathcal{H}_N$, $i=1,2,\dots N$ and $\rho_i$'s are the eigenvalues of $\rho$. The property $(\rm{iv}) \rm{Tr}\rho^2\leq 1$ follows from Eq. \eqref{eq:rho_prop}. Equality holds when $\rho$ is a pure state. 

Following these properties, the operators $\hat{x}_i$ satisfy
\begin{align*}
    ({\rm i}) \hat{x}_i = \hat{x}_i^\dagger \quad ({\rm ii}) \text{Tr} [\hat{x}_i] = 0 \quad ({\rm iii}) \text{Tr} [\hat{x}_i\hat{x}_j] = 2\delta_{ij}.
\end{align*}
The $x_i$'s are characterised with structure constants $f_{ijk}$, completely asymmetric tensor and $g_{ijk}$, completely symmetric tensor of Lie algebra
\begin{alignat}{1}
    [\hat{x}_i,\hat{x}_j] &= 2if_{ijk} \hat{x}_k\\
    \{\hat{x}_i,\hat{x}_j\}&=\frac{2}{N}\delta_{ij}\mathbb{\hat{I}}_N+2g_{ijk}\hat{x}_k.
\end{alignat}
By imposing (iv), the length of the operators $\hat{x}_i$ are restricted as 
\begin{equation}
    |x|\equiv \sqrt{x_ix_j} \leq \sqrt{\frac{2(N-1)}{N}}.
\end{equation}
Systematic construction of the generators generalising the Pauli spin operators for an $N$-level system is given by \cite{hioe, pottinger_lendi, lendi}
\begin{alignat}{1}
    \{\hat{x}_i\}_{i=1}^{N^2-1} = \{\hat{u}_{jk},\hat{v}_{jk},\hat{w}_l\}
\end{alignat}
where
\begin{alignat}{1}\label{eq:op_rho}
    \hat{u}_{jk} &= \ket{j}\bra{k} + \ket{k} \bra{j},\\
    \hat{v}_{jk} &= -\iota(\ket{j}\bra{k} - \ket{k}\bra{j}),\\
    \hat{w}_l &= \sqrt{\frac{2}{l(l+1)}}\left( \sum_{j=1}^l \ket{j}\bra{j}-l\ket{l+1}\bra{l+1}\right),\\
    &1\leq j < k \leq N, 1\leq l \leq N-1.
\end{alignat}
For $N=2$,
\begin{alignat}{1}
    \hat{x}_1 &= \hat{u}_{12} = \ket{1}\bra{2} + \ket{2} \bra{1} \equiv \hat{X},\nonumber\\
    \hat{x}_2 &= \hat{v}_{12} = -\iota (\ket{1}\bra{2} - \ket{2}\bra{1}) \equiv \hat{Y},\nonumber\\
    \hat{x}_3 &= \hat{w}_l= \ket{1}\bra{1} - \ket{2}\bra{2} \equiv \hat{Z},
\end{alignat}
where $\ket{1}=\begin{bmatrix}
    1 & 0 \\
\end{bmatrix}^{\rm T}$ and $\ket{2}=\begin{bmatrix}
    0 & 1 \\
\end{bmatrix}^{\rm T}$ and the structure constants are $f_{ijk} = \epsilon_{ijk}$ (Levi-Civita), $g_{ijk}=0$.


\newpage
\begin{thebibliography}{99}

\bibitem{ms} B. Misra and E. C. G. Sudarshan, J. Math. Phys. {\bf 18}, 756 (1977).

\bibitem{dehmelt} H. G. Dehmelt, Bull. Am. Phys. Soc. {\bf 20}, 60 (1975).

\bibitem{deh} H. G. Dehmelt, IEEE Transactions on Instrumentation and Measurement, IM{\bf 31},  83 (1982).

\bibitem{minev} Z. Minev et al., Nature {\bf 570}, 200 (2019).

\bibitem{parveen} K. Snizhko, P. Kumar, A. Romito, Phys. Rev. Res. {\bf 2}, 033512 (2020).

\bibitem{krjj} Komal Kumari, Garima Rajpoot, Sandeep Joshi, and Sudhir R. Jain, Ann. Phys. {\bf 450}, 169222 (2023). 

\bibitem{kimura} G. Kimura, \textit{The Bloch vector for N-level systems}, Phys. Lett. A {\bf 314}, 339 (2003).

\bibitem{jordan} A. Chantasri, J. Dressel, A. Jordan,  Phys. Rev. A {\bf 88}, 042110 (2013).

\bibitem{jordan2} A. Chantasri, A. Jordan,  Phys. Rev. A.{\bf 92}, 032125 (2015).

\bibitem{rieffel} E. Rieffel, Wolfgang Polak, \textit{Quantum Computing: A gentle introduction}, (The MIT Press, Cambridge) (2011).

\bibitem{hioe} F. T. Hioe and J. H. Eberly, \textit{$N$-Level Coherence Vector and Higher Conservation Laws in Quantum Optics and Quantum Mechanics}, Phys. Rev. Lett. \textbf{47}, 838 (1981).

\bibitem{pottinger_lendi} J. P\"{o}ttinger and K. Lendi, \textit{Generalized Bloch equations for decaying systems}, Phys. Rev. A {31}, 1299 (1985).

\bibitem{lendi} K. Lendi, \textit{Entropy production in coherence-vector formulation for $N$-level systems}, Phys. Rev. A {34}, 662 (1986).


\end{thebibliography}
\end{document}